\documentclass[aps, prd, floatfix, nofootinbib, superscriptaddress, twocolumn]{revtex4-1}

\usepackage{latexsym}
\usepackage{amsmath}
\usepackage{amssymb}
\usepackage{amsfonts}

\usepackage[mathscr,scaled=1.15]{urwchancal}
\DeclareFontFamily{OT1}{pzc}{}
\DeclareFontShape{OT1}{pzc}{m}{it}%
{<-> s * [1.15] pzcmi7t}{}
\DeclareMathAlphabet{\mathpzc}{OT1}{pzc}{m}{it}

\usepackage{color}

\usepackage{supertabular}
\usepackage{placeins}
\usepackage{epsfig}
\usepackage{graphicx}

\definecolor{purple}{rgb}{0.5,0,0.5}
\definecolor{blue}{rgb}{0.0,0,0.9}
\definecolor{prdblue}{rgb}{0.133,0.118,0.498}
\usepackage[colorlinks=true, pdfstartview=FitV, linkcolor=prdblue, citecolor= prdblue, urlcolor=prdblue]{hyperref}

\begin{document}


\title{Spectrum and structure of octet and decuplet baryons\\ and their positive-parity excitations}



\author{Chen Chen}
\email[]{Current address: Institut f\"ur Theoretische Physik, Justus-Liebig-Universit\"at Gie{\ss}en, 35392 Gie{\ss}en, Germany\\ Chen.Chen@theo.physik.uni-giessen.de}
\affiliation{Instituto de F\'isica Te\'orica, Universidade Estadual Paulista, Rua Dr.~Bento Teobaldo Ferraz, 271, 01140-070 S\~ao Paulo, SP, Brazil}

\author{Gast\~ao Krein}
\affiliation{Instituto de F\'isica Te\'orica, Universidade Estadual Paulista, Rua Dr.~Bento Teobaldo Ferraz, 271, 01140-070 S\~ao Paulo, SP, Brazil}

\author{Craig D. Roberts}
\email[]{cdroberts@anl.gov}
\affiliation{Physics Division, Argonne National Laboratory, Lemont, Illinois
60439, USA}

\author{Sebastian M. Schmidt}
\affiliation{
Institute for Advanced Simulation, Forschungszentrum J\"ulich and JARA, D-52425 J\"ulich, Germany}

\author{Jorge Segovia}
\affiliation{Departamento de Sistemas F\'{\i}sicos, Qu\'{\i}micos y Naturales,
Universidad Pablo de Olavide, E-41013 Sevilla, Spain}

\date{15 July 2019}

\begin{abstract}
A continuum approach to the three valence-quark bound-state problem in quantum field theory, employing parametrisations of the necessary kernel elements, is used to compute the spectrum and Poincar\'e-covariant wave functions for all flavour-$SU(3)$ octet and decuplet baryons and their first positive-parity excitations.  Such analyses predict the existence of nonpointlike, dynamical quark-quark (diquark) correlations within all baryons; and a uniformly sound description of the systems studied is obtained by retaining flavour-antitriplet--scalar and flavour-sextet--pseudovector diquarks.  Thus constituted, the rest-frame wave function of every system studied is primarily $S$-wave in character; and the first positive-parity excitation of each octet or decuplet baryon exhibits the characteristics of a radial excitation. Importantly, every ground-state octet and decuplet baryon possesses a radial excitation.  Hence, the analysis predicts the existence of positive-parity excitations of the $\Xi$, $\Xi^\ast$, $\Omega$ baryons, with masses, respectively (in GeV): 1.84(08), 1.89(04), 2.05(02).  These states have not yet been empirically identified.  This body of analysis suggests that the expression of emergent mass generation is the same in all $u$, $d$, $s$ baryons and, notably, that dynamical quark-quark correlations play an essential role in the structure of each one.  It also provides the basis for developing an array of predictions that can be tested in new generation experiments.
\end{abstract}



\maketitle


\section{Introduction}\label{introduction}
%
The nucleon's first excited state, the $\Delta$-baryon, was discovered almost 70 years ago \cite{Anderson:1952nw}.  In the intervening period, baryon spectroscopy has, \emph{inter alia}: delivered the quark model \cite{GellMann:1964nj, Zweig:1981pd} and the colour quantum number \cite{Greenberg:1964pe}; and played a crucial role in the development of quantum chromodynamics (QCD) \cite{Marciano:1977su, Marciano:1979wa}.  Quark potential models based on three flavours of light quark -- $u$, $d$, $s$ \cite{Capstick:2000qj, Crede:2013sze, Giannini:2015zia} can describe the known spectrum of ground-state flavour-$SU(3)$ octet and decuplet baryons \cite{Tanabashi:2018oca}.  They also predict many related excited states; a feature which initiated a widespread international search for such systems \cite{Aznauryan:2012baS, Briscoe:2015qia, Mokeev:2018zxt, Burkert:2018oyl, Carman:2018fsn, Thiel:2018eif, Sokhoyan:2018geu, Cole:2018faq, Suzuki:2009nj, Suzuki:2010yn, Anisovich:2011fc, Ronchen:2012eg, Kamano:2013iva, Kamano:2015hxa, Landay:2018wgf}.  The search is still underway because only a small percentage of the predicted states have thus far been found.  This is the ``missing (baryon) resonance'' problem.

One solution to the missing resonance problem is to suppose that typical three-quark potential models do not provide a sound guide to baryon excitations.  An alternative is to reduce the number of degrees-of-freedom within the system by describing baryons as two-body quark$+$pointlike-diquark bound-states \cite{Lichtenberg:1968zz, Santopinto:2016zkl}, in which case the number of excited states is much diminished.  However, given that the baryon spectrum computed using lattice-regularised QCD (lQCD) possesses a number of states that is consistent with three-quark models \cite{Edwards:2011jj}, the pointlike-diquark answer appears incorrect.  All the same, it does not follow that the diquark concept is eliminated: only that such quark-quark correlations cannot be pointlike.

\begin{figure}[b]
\centerline{%
\includegraphics[clip, width=0.45\textwidth]{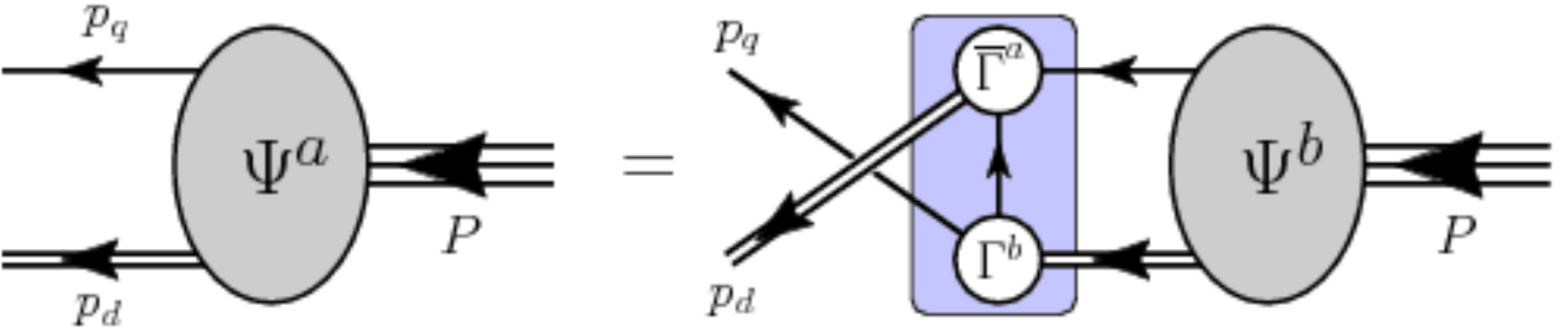}}
\caption{\label{figFaddeev}
Poincar\'e covariant Faddeev equation: a linear integral equation for the matrix-valued function $\Psi$, being the Faddeev amplitude for a baryon of total momentum $P= p_q + p_d$, which expresses the relative momentum correlation between the dressed-quarks and -nonpointlike-diquarks within the baryon.  The shaded rectangle demarcates the kernel of the Faddeev equation:
\emph{single line}, dressed-quark propagator (Sec.\,\ref{SecDressedq}); $\Gamma$,  diquark correlation amplitude (Sec.\,\ref{SecGammaqq}); and \emph{double line}, diquark propagator (Sec.\,\ref{SecPropsqq}). }
\end{figure}

The analysis of a baryon as a three--valence-body bound-state problem in continuum quantum field theory became possible following the formulation of a Poincar\'e-covariant Faddeev equation \cite{Cahill:1988dx, Burden:1988dt, Cahill:1988zi, Reinhardt:1989rw, Efimov:1990uz}, which is depicted in Fig.\,\ref{figFaddeev}.  This approach assumes that dynamical, nonpointlike diquark correlations play an important role in baryon structure.  It is founded on a simplified treatment of the scattering problem in the two-body quark-quark subchannels of the full three-body problem (see, \emph{e.g}.\ Ref.\,\cite{Hecht:2002ej}, Sec.\,II.A.2), which capitalises on the observation that the same interaction which describes colour-singlet mesons also generates diquark correlations in the colour-antitriplet $(\bar 3)$ channel \cite{Cahill:1987qr, Maris:2002yu}.  Whilst the diquarks do not survive as asymptotic states, \emph{viz}.\ they do not appear in the strong interaction spectrum \cite{Bender:1996bb, Bhagwat:2004hn}, the attraction between the quarks in the $\bar 3$ channel draws a picture in which two quarks are always correlated as a colour-$\bar 3$ diquark pseudoparticle, and binding is effected by the iterated exchange of roles between the bystander and diquark-participant quarks.  This approach to the spectrum and interactions of baryons has been applied widely with phenomenological success, \emph{e.g}.\, Refs.\,\cite{Segovia:2014aza, Segovia:2015hra, Segovia:2015ufa, Segovia:2016zyc, Eichmann:2016hgl, Eichmann:2016nsu, Lu:2017cln, Chen:2017pse, Mezrag:2017znp, Chen:2018nsg}.

A spectrum of flavour-$SU(3)$ octet and decuplet baryons, their parity partners, and the radial excitations of these systems, was computed in Ref.\,\cite{Lu:2017cln} using a symmetry-preserving treatment of a vector$\,\times\,$vector contact-interaction (SPCI) as the foundation for the relevant few-body equations.  %
Namely, in the gap- and Bethe-Salpeter equations solved to prepare the baryon bound-state kernel, a momentum-independent vector-boson-exchange kernel was used, with the ultraviolet divergence regulated so as to ensure the relevant Ward-Green-Takahashi identities are preserved \cite{GutierrezGuerrero:2010md}.
With such an approach: the diquark bound-state amplitudes are momentum-independent but the diquarks are dynamical degrees-of-freedom with nonzero electromagnetic radii \cite{Roberts:2011wy}; and one obtains a spectrum that is qualitatively equivalent to the quark-model and lQCD results.

Realistic momentum-dependent dynamics was included in Ref.\,\cite{Eichmann:2016hgl}, which computes a reduced flavour-$SU(2)$ spectrum that qualitatively matches the analogous quark-model result.  Aspects of that formulation, however, prevent direct access to the mass-shell for many of the predicted states, hindering computation of the associated wave functions and hence limiting the available structural information.

Herein, we avoid such difficulties by employing a QCD-kindred framework, used elsewhere \cite{Chen:2017pse} to perform a comparative study of the four lightest $(I=1/2,J^P = 1/2^\pm)$ baryon isospin-doublets; extending it to flavour-$SU(3)$ and delivering therewith a spectrum of octet and decuplet baryons and their first positive-parity excitations along with structural insights drawn from analyses of their Poincar\'e-covariant wave functions.  Our results should supply the foundation for subsequent calculations of a diverse array of baryon-resonance electroproduction form factors, existing empirical information on which \cite{Armstrong:1998wg, Thompson:2000by, Aznauryan:2004jd, Denizli:2007tq, Dalton:2008aa, Aznauryan:2008pe, Aznauryan:2009mx, Dugger:2009pn, Aznauryan:2011qj, Mokeev:2012vsa, Mokeev:2013kka, Mokeev:2015lda,Aznauryan:2012ec, Aznauryan:2012ba,  Park:2014yea, Burkert:2016dxc, Isupov:2017lnd, Fedotov:2018oan} along with recent photocoupling measurements \cite{Golovatch:2018hjk}, will be enlarged by forthcoming experiments at the Thomas Jefferson National Accelerator Facility (JLab).

We emphasise that the continuum analyses indicated above form part of the body of Dyson-Schwinger equation (DSE) studies of hadron structure \cite{Roberts:1994dr, Roberts:2000aa, Maris:2003vk, Roberts:2007jh, Boucaud:2011ug, Roberts:2015lja, Aguilar:2015bud, Pennington:2016dpj, Horn:2016rip, Mezrag:2016hnp, Eichmann:2016yit, Fischer:2018sdj}.
%
%
In this approach, the challenge is a need to employ a truncation so as to define a tractable problem.  Much has been learnt; and one may now separate DSE studies into three classes.
\emph{Class-A}.\ model-independent statements about QCD;
\emph{Class-B}.\ illustrations of such statements using well-constrained model elements and possessing a traceable connection to QCD;
\emph{Class-C}.\ QCD-inspired analyses whose elements have not been computed using a truncation that preserves a systematically-improvable connection with QCD.
The analysis described herein lies within Class-C.

We describe our approach to the baryon bound-state problem in Sec.\,\ref{secFaddeev}; and detail and explain the character of the solutions for the octet and decuplet baryons and their first positive-parity excitations in Sec.\,\ref{solutions}.  Section\,\ref{epilogue} provides a summary and indicates some new directions.

\section{Baryon Bound State Problem}
\label{secFaddeev}
\subsection{Faddeev equation}
In its general form, the Faddeev equation sums all possible exchanges and interactions that can take place between the three dressed-quarks that express a baryon's valence-quark content.  Used with a realistic quark-quark interaction \cite{Hopfer:2013np, Binosi:2014aea, Williams:2015cvx, Binosi:2016wcx, Binosi:2016nme, Cyrol:2017ewj, Rodriguez-Quintero:2018wma}, it predicts the appearance of soft (nonpointlike) fully-interacting diquark correlations within baryons, whose characteristics are greatly influenced by dynamical chiral symmetry breaking (DCSB) \cite{Segovia:2015ufa}.  Consequently, the problem of determining a baryon's mass, internal structure, etc., is transformed into that of solving the linear, homogeneous matrix equation depicted in Fig.\,\ref{figFaddeev}.

\subsection{Dressed quarks}
\label{SecDressedq}
Regarding flavour-$SU(3)$ octet and decuplet baryons and their radial excitations, the Faddeev kernel in Fig.\,\ref{figFaddeev} involves three basic elements, \emph{viz}.\ the dressed light-quark propagators, $S_f(p)$, $f=u,d,s$, and the correlation amplitudes and propagators for all participating diquarks.
Much is known about $S_f(p)$, and in constructing the kernel we use the algebraic forms described in Appendix~\ref{appendixKFE}, which have proven efficacious in the explanation and unification of a wide range of hadron observables \cite{Ivanov:1998ms, Ivanov:2007cw, Segovia:2014aza, Segovia:2015hra,Segovia:2015ufa, Segovia:2016zyc}.
(\emph{N.B}.\ We assume isospin symmetry throughout, \emph{i.e}.\ $u$- and $d$-quarks are mass-degenerate and described by the same propagator.  Consequently, all diquarks in an isospin multiplet are degenerate.)

\subsection{Correlation amplitudes}
\label{SecGammaqq}
In Fig.\,\ref{figFaddeev}, all participating diquarks are colour-antitriplets because they must combine with the bystander quark to form a colour singlet.  Notably, the colour-sextet quark+quark channel does not support correlations because gluon exchange is repulsive in this channel \cite{Cahill:1987qr}.

Diquark isospin-spin structure is more complex.  Accounting for Fermi-Dirac statistics, five types of correlation are possible in a $J=1/2$ bound-state: flavour-$\bar 3$--scalar, flavour-$6$--pseudo\-vector, flavour-$\bar 3$--pseudo\-scalar, flavour-$\bar 3$--vector, and flavour-$6$--vector.  However, only the first two are important in positive-parity systems \cite{Eichmann:2016hgl, Lu:2017cln, Chen:2017pse}; and the associated leading correlation amplitudes are, respectively:
{\allowdisplaybreaks
\begin{subequations}
\label{qqBSAs}
\begin{align}
\Gamma_{0^+}^j(k;K) & = g_{0^+}^j\,\vec{H} \,T_{\bar 3_f}^j \, \gamma_5 C\, {\mathpzc F}(k^2/[\omega_{0^+}^j]^2) \,, \\
%
%
\Gamma_{1^+ \mu}^{g}(k;K)
    & = i g_{1^+}^g \, \vec{H} \, T_{6_f}^g \, \gamma_\mu C \, {\mathpzc F}(k^2/[\omega_{1^+}^g]^2)\,, \label{flavour6}
\end{align}
\end{subequations}}
\hspace*{-0.5\parindent}where:
$K$ is the total momentum of the correlation,
$k$ is a two-body relative momentum,
${\mathpzc F}$ is the function in Eq.\,\eqref{defcalF},
$\omega_{J^P}^{j,g}$ are size parameters, and $g_{J^P}^{j,g}$ are couplings into the channel, fixed by normalisation;
$\vec{H} = \{i\lambda_c^7, -i\lambda_c^5,i\lambda_c^2\}$, with $\{\lambda_c^k,k=1,\ldots,8\}$ denoting Gell-Mann matrices in colour-space, expresses the diquarks' colour antitriplet character;
$C=\gamma_2\gamma_4$ is the charge-conjugation matrix;
\begin{subequations}
\label{Tmatrices}
\begin{align}
\{T_{\bar 3_f}^j,j& =1,2,3\} = \{i\lambda^2,i\lambda^5,i\lambda^7\}\,, \\
\nonumber
\{T_{6_f}^g,g & =1,\ldots,6\}  =\{
s_0 \lambda^0 +s_3 \lambda^3  + s_8 \lambda^8 ,
\lambda^1,
\lambda^4, \\
&  s_0 \lambda^0 - s_3 \lambda^3  + s_8 \lambda^8,
\lambda^6,
s_0 \lambda^0 - 2 s_8 \lambda^8
\}\,,
\end{align}
\end{subequations}
with $s_0=\surd 2/3$, $s_3=1/\surd 2$, $s_8=1/\surd 6$,
$\{\lambda^k,k=1,\ldots,8\}$ denoting flavour-$SU(3)$ Gell-Mann matrices,
$\lambda^0={\rm diag}[1,1,1]$, and
all flavour matrices left-active on column$[u,d,s]$.
(Our Euclidean metric conventions are explained in Ref.\,\cite{Segovia:2014aza}, Appendix\,B.)

Turning to decuplet baryons, since it is not possible to combine a $\bar 3_f$ diquark with a $3_f$-quark to obtain a member of the symmetric $10_f$ representation of $SU(3)_f$, decuplet baryons only contain $6_f$--axial-vector diquarks, which are associated with the amplitudes in Eq.\,\eqref{flavour6}.

The amplitudes in Eqs.\,\eqref{qqBSAs} are normalised canonically:
\begin{subequations}
\label{CanNorm}
\begin{align}
2 K_\mu & = \left. \frac{\partial }{\partial Q_\mu} \,  \Pi(K;Q)\right|_{Q=K}^{K^2 = - [m_{J^P}^{j,g}]^{2}},\\
\nonumber
\Pi(K;Q) & = {\rm tr} \int \frac{d^4 k}{(2\pi)^4} \bar \Gamma(k;-K) {\mathpzc S}(k+Q/2) \\
& \quad \times \Gamma(k;K) {\mathpzc S}^{\rm T}(-k+Q/2)\,, \label{eqPi}
\end{align}
\end{subequations}
where $\bar\Gamma(k;K) = C^\dagger \Gamma(-k;K) C$,
$[\cdot]^{\rm T}$ denotes matrix transpose,
and ${\mathpzc S} = {\rm diag}[S_u,S_d,S_s]$.
When the involved correlation amplitudes carry Lorentz indices $\mu$, $\nu$, the left-hand-side of Eq.\,\eqref{eqPi} also includes a factor $\delta_{\mu\nu}$.  It is apparent now that the strength of coupling in each channel, $g_{J^P}^{j,g}$ in Eq.\,\eqref{qqBSAs}, is fixed by the associated value of $\omega_{J^P}^{j,g}$.

\subsection{Diquark propagators, masses, couplings}
\label{SecPropsqq}
A propagator is associated with each quark-quark correlation in Fig.\,\ref{figFaddeev}; and we use \cite{Segovia:2014aza, Chen:2017pse}:
\begin{subequations}
\label{Eqqqprop}
\begin{align}
\Delta^{0^+j }(K) & = \frac{1}{[m^j_{0^+}]^2} \, {\mathpzc F}(K^2/[\omega^j_{0^+}]^2)\,,\\
\Delta^{1^+g}_{\mu\nu}(K) & = \left[ \delta_{\mu\nu} + \frac{K_\mu K_\nu}{[m^g_{1^+}]^2} \right]
 \frac{1}{[m^g_{1^+}]^2} \, {\mathpzc F}(K^2/[\omega^g_{1^+}]^2)\,.
\end{align}
\end{subequations}
These algebraic forms ensure that the diquarks are confined within the baryons, as appropriate for coloured correlations: whilst the propagators are free-particle-like at spacelike momenta, they are pole-free on the timelike axis.  This is sufficient to ensure confinement via the violation of reflection positivity (see, e.g.\ Ref.\,\cite{Horn:2016rip}, Sec.\,3).

The diquark masses and widths are related via
\begin{equation}
m^{j,g}_{J^P} = \surd 2 \, \omega^{j,g}_{J^P}.
\end{equation}
This identification accentuates the free-particle-like propagation characteristics of the diquarks within the baryon \cite{Segovia:2014aza}.  The mass-scales are constrained by numerous studies; and we use (in GeV):
\begin{subequations}
\label{dqmasses}
\begin{align}
\label{dqmassessc}
m^{(1)}_{0^+} & = 0.80\,,\;
m^{(2,3)}_{0^+} = 0.95\,, \\
\label{dqmassesax}
m^{(4,5,7)}_{1^+} & = 0.90\,,\;  
m^{(6,8)}_{1^+} = 1.05\,,\;
m^{(9)}_{1^+} = 1.20\,,
\end{align}
\end{subequations}
where the values of $m^{(j=1)}_{0^+}$ and $m^{(g=4,5,7)}_{1^+}$ are drawn from Refs.\,\cite{Segovia:2014aza, Segovia:2015hra}, because they provide for a good description of numerous dynamical properties of the nucleon, $\Delta$-baryon and Roper resonance; and the masses $m^{(j=2,3)}_{0^+}$, $m^{(g=6,8)}_{1^+}$ and $m^{(g=9)}_{1^+}$ are derived therefrom via an equal-spacing rule, \emph{viz}.\ replacing one light-quark by a $s$-quark brings an extra $0.15\,{\rm GeV} \approx M_s^E - M_u^E$ from Eq.\,\eqref{MEq}.  This is the standard response found in solutions of the relevant Bethe-Salpeter equations \cite{Qin:2018dqp}; and for completeness, we will subsequently display results with all diquark masses varied by $\pm 5$\%.

Using Eqs.\,\eqref{qqBSAs}, \eqref{CanNorm}, and the masses in Eqs.\,\eqref{dqmasses}:
\begin{subequations}
\label{dqcouplings}
\begin{align}
\label{dqcouplingssc}
g^{(1)}_{0^+} & = 14.75\,,\;
g^{(2,3)}_{0^+} = 9.45\,,\\
\label{dqcouplingsax}
g^{(4,5,7)}_{1^+} & = 12.73\,,\;
g^{(6,8)}_{1^+} = 7.62\,,\;
g^{(9)}_{1^+} = 3.72\,.
\end{align}
\end{subequations}
Given that it is the coupling-squared which appears in the Faddeev kernels, scalar diquarks will dominate the Faddeev amplitudes of $J=1/2$ baryons; but pseudovector diquarks must also play a material role because $g^2_{1^+}/g^2_{0^+} \approx 0.7$.

\subsection{Remarks on the Faddeev kernels}
\label{FEremarks}
The elements described in the preceding subsections are sufficient to specify the Faddeev kernels describing the quark cores of all positive-parity octet and decuplet baryons.  Moreover, owing to our deliberate use of algebraic parametrisations for these inputs, the Faddeev equations thus obtained can be solved directly on the baryon mass-shells, providing simultaneously the associated on-shell Faddeev amplitudes and wave functions.

It should be recorded, however, that whilst the algebraic forms we use are based on existing calculations and associated insights \cite{Ivanov:1998ms, Ivanov:2007cw}, their pointwise forms are not necessarily accurate representations of QCD's solutions for these quantities on the complete momentum domains available to their arguments.  Efforts to determine such pointwise accurate representations are ongoing, \emph{e.g}.\, Refs.\,\cite{Strauss:2012dg, Windisch:2016iud, Lowdon:2017gpp, Cyrol:2018xeq, Siringo:2018uho, Hayashi:2018giz, Dudal:2019gvn, Binosi:2019ecz}.  Nevertheless, on the domains sampled in solving the Faddeev equations, our algebraic forms are fair approximations, as highlighted, \emph{e.g}.\ in Ref.\,\cite{Chen:2017pse}, Appendix~A; and this is sufficient for our purposes.

Usefully, too, the inputs we use for the propagators and correlation amplitudes are constrained by observables and hence they express many effects that are lost in straightforward implementations of the lowest-order (rainbow-ladder, RL) truncation of the bound-state equations \cite{Binosi:2016rxz}.

Notwithstanding that, some correction of the Faddeev kernels is necessary to overcome an intrinsic weakness of the equation depicted in Fig.\,\ref{figFaddeev}. 
Namely, resonant contributions, \emph{viz}.\ meson-baryon final-state-interactions (MB\,FSIs), should be included \cite{Eichmann:2008ae, Eichmann:2008ef}.

It is essential not to miscount when incorporating these effects.  In practical calculations they divide into two distinct types.  The first is within the gap equation, where pseudoscalar-meson loop-corrections to the dressed-gluon-quark vertex act to reduce uniformly the mass-function of a dressed-quark \cite{Eichmann:2008ae, Fischer:2008wy, Eichmann:2008ef, Cloet:2008fw}.  This effect can be pictured as a single quark emitting and reabsorbing a pseudoscalar meson.  It can be mocked-up by simply choosing the parameters in the gap equation's kernel so as to obtain dressed-quark mass-functions that are characterised by mass-scales $M_u^E  \approx 0.4\,$GeV, $M_s^E \approx 0.5\,$GeV.  Such an approach has been widely employed with phenomenological success \cite{Roberts:2000aa, Maris:2003vk, Roberts:2007jh, Roberts:2015lja, Eichmann:2016yit}, and is implicit herein.

The second type of correction arises in connection with bound-states and may be likened to adding pseudoscalar meson exchange \emph{between} dressed-quarks within the bound-state \cite{Hollenberg:1992nj, Alkofer:1993gu, Ishii:1998tw, Pichowsky:1999mu, Hecht:2002ej}, as opposed to the first type of effect, \emph{i.e}.\ emission and absorption of a meson by the \emph{same} quark.  The type-2 contribution is that computed in typical evaluations of meson-loop corrections to hadron observables based on a point-hadron Lagrangian \cite{Thomas:1999mu}.  This fact should be borne in mind when estimating the size of meson-loop corrections to the quark-core masses of baryons computed using the equation depicted in Fig.\,\ref{figFaddeev}.

\subsection{Faddeev amplitudes}
In solving the Faddeev equation, Fig.\,\ref{figFaddeev}, one obtains both the mass-squared and bound-state amplitude of all baryons with a given value of $J^P$.  In fact, it is the form of the Faddeev amplitude which fixes the channel.  A baryon is described by
\begin{align}
\Psi^B & = \psi^B_1 + \psi^B_2 + \psi^B_3\,,
\end{align}
where the subscript identifies the bystander quark, \emph{i.e}.\ the quark that is not participating in a diquark correlation,  $\psi^B_{1,2}$ are obtained from $\psi^B_3=:\psi^B$ by a cyclic permutation of all quark labels.

For an octet baryon ($B_{\underline 8}=N,\Lambda,\Sigma,\Xi$; $J^P=1/2^+$),
\begin{align}
\nonumber
& \psi^{\underline 8}(p_i,\alpha_i,\sigma_i) \\
\nonumber
 = & \sum_{j\in B_{\underline 8}} \, [\Gamma^j_{0^+}(k;K)]^{\alpha_1 \alpha_2}_{\sigma_1 \sigma_2} \, \Delta^{0^+ j}(K) \,[\varphi_{0^+}^{{\underline 8} j}(\ell;P) u(P)]^{\alpha_3}_{\sigma_3} \\
  + &  \sum_{g\in B_{ \underline 8}} \, [\Gamma^g_{1^+\mu}]^{\alpha_1 \alpha_2}_{\sigma_1 \sigma_2} \, \Delta^{1^+ g}_{\mu\nu} \, [\varphi_{1^+\nu }^{{\underline 8} g}(\ell;P) u(P)]^{\alpha_3}_{\sigma_3} \,,
\label{FaddeevAmp8}
\end{align}
where
$(p_i,\sigma_i,\alpha_i)$ are the momentum, spin and isospin labels of the quarks constituting the bound state;
$P=p_1 + p_2 + p_3=p_d+p_q$ is the total momentum of the baryon;
$k=(p_1-p_2)/2$, $K=p_1+p_2=p_d$, $\ell = (-K + 2 p_3)/3$;
$j$ and $g$ are the labels in Eqs.\,\eqref{Tmatrices} and the sums run over those flavour-combinations permitted in $B_{\underline 8}$, detailed in Appendix~\ref{appendixAssorted};
and $u(P)$ is a Euclidean spinor (see Ref.\,\cite{Segovia:2014aza}, Appendix\,B for details).
The remaining elements in Eq.\,\eqref{FaddeevAmp8} are the following matrix-valued functions:
\begin{subequations}
\label{8sa}
\begin{align}
\varphi_{0^+}^{{\underline 8} j}(\ell;P) & = \sum_{k=1}^2 {\mathpzc s}_{{\underline 8} k}^{ j}(\ell^2,\ell\cdot P)\,  {\mathpzc S}^k(\ell;P) \,, \\
\varphi_{1^+ \nu}^{{\underline 8} g}(\ell;P)  & = \sum_{k=1}^6 {\mathpzc a}_{{\underline 8} k}^{ g+3}(\ell^2,\ell\cdot P)\, \gamma_5 {\mathpzc A}^k_\nu(\ell;P) \,,
\end{align}
\end{subequations}
where
\begin{align}
\label{diracbasis}
\nonumber
{\mathpzc S}^1 & = {\mathbf I}_{\rm D} \,,\;
{\mathpzc S}^2  = i \gamma\cdot\hat\ell - \hat\ell \cdot\hat P {\mathbf I}_{\rm D}\,, \\
{\mathpzc A}^1_\nu & =  \gamma\cdot\ell^\perp \hat P_\nu\,,\;
{\mathpzc A}^2_\nu  = - i \hat P_\nu {\mathbf I}_{\rm D}\,,\;
{\mathpzc A}^3_\nu  = \gamma\cdot\hat\ell^\perp \hat\ell^\perp_\nu\,, \\
\nonumber
{\mathpzc A}^4_\nu & = i\hat \ell_\nu^\perp {\mathbf I}_{\rm D}\,,\;
{\mathpzc A}^5_\nu = \gamma_\nu^\perp - {\mathpzc A}^3_\nu\,,\;
{\mathpzc A}^6_\nu = i \gamma_\nu^\perp \gamma\cdot\hat\ell^\perp - {\mathpzc A}^4_\nu\,,
\end{align}
with $\hat\ell^2=1$, $\hat P^2 = -1$, $\ell^\perp = \hat\ell_\nu +\hat\ell\cdot\hat P \hat P_\nu$, $\gamma^\perp = \gamma_\nu +\gamma\cdot\hat P \hat P_\nu$.

Owing to the symmetry-prescribed absence of flavour-$\bar 3$ components, the structure of decuplet baryons is simpler ($B_{\underline{10}} = \Delta, \Sigma^\ast, \Xi^\ast, \Omega$; $J^P=3/2^+$):
\begin{align}
\nonumber
& \psi_\mu^{\underline{10}}(p_i,\alpha_i,\sigma_i) \\
 = &  \sum_{g\in B_{\underline{10}}} \, [\Gamma^g_{1^+\mu}]^{\alpha_1 \alpha_2}_{\sigma_1 \sigma_2} \, \Delta^{1^+ g}_{\mu\nu} \, [\varphi_{\nu\rho}^{\underline{10} g}(\ell;P) u_\rho(P)]^{\alpha_3}_{\sigma_3} \,,
\label{FaddeevAmp10}
\end{align}
where $u_\rho(P)$ is a Rarita-Schwinger spinor (Ref.\,\cite{Segovia:2014aza}, Appendix\,B); and, with ${\mathpzc S}^k$ and ${\mathpzc A}_\nu^k$ in Eq.\,\eqref{diracbasis},
\begin{subequations}
\label{10d}
\begin{align}
\varphi_{\nu\rho}^{\underline{10} g}(\ell;P) & =
\sum_{k=1}^8 {\mathpzc a}_{\underline{10} k}^{g+3}(\ell^2,\ell\cdot P)\, {\mathpzc D}^k_{\nu\rho}(\ell;P)\,,\\
& {\mathpzc D}^k_{\nu\rho} = {\mathpzc S}^k\,\delta_{\nu\rho}\,,   \rule{8ex}{0ex} k=1,2\,,\\
& {\mathpzc D}^k_{\nu\rho} = i\gamma_5\,{\mathpzc A}_\nu^{k-2}\,\ell^\perp_\rho\,, \quad k=3,\dots,8\,.
\end{align}
\end{subequations}

Inserting the appropriate amplitude into the equation defined by Fig.\,\ref{figFaddeev},  the specific form of the bound-state equation for the considered baryon is obtained.  We do not record these equations herein. They are readily derived following the procedures detailed, \emph{e.g}.\ in Ref.\,\cite{Segovia:2014aza}.

\subsection{Faddeev wave functions}
\label{SecFWFs}
The (unamputated) Faddeev wave function can be computed from the amplitude specified by Eqs.\,\eqref{FaddeevAmp8}\,--\,\eqref{10d}
%
simply by attaching the appropriate dressed-quark and diquark propagators.  It may also be decomposed in the form of Eqs.\,\eqref{8sa}, \eqref{10d}.  Naturally, the scalar functions are different, and we label them $\tilde{\mathpzc s}_{\underline{8} k}^{j}$, $\tilde{\mathpzc a}_{\underline{8} k}^{g}$, $\tilde {\mathpzc a}_{\underline{10} k}^{g}$.

Both the Faddeev amplitude and wave function are Poincar\'e covariant, \emph{i.e}.\ they are qualitatively identical in all reference frames.  Naturally, each of the scalar functions that appears is frame-independent, but the frame chosen determines just how the elements should be combined.  Consequently, the manner by which the dressed-quarks' spin, $S$, and orbital angular momentum, $L$, add to form a particular $J^P$ combination is frame-dependent: $L$, $S$ are not independently Poincar\'e invariant.\footnote{The nature of the combination is also scale dependent because the definition of a dressed-quark and the character of the correlation amplitudes changes with resolving scale, $\zeta$, in a well-defined manner \cite{Lepage:1980fj, Raya:2015gva}.  Our analysis is understood to be valid at $\zeta \simeq 1\,$GeV.}
Hence, in order to enable comparisons with typical formulations of constituent quark models, here we subsequently list the set of baryon rest-frame quark-diquark angular momentum identifications \cite{Oettel:1998bk, Cloet:2007piS}.

\begin{table*}[t]
\caption{\label{OctetDecupletMasses}
Computed dressed-quark-core masses of ground-state octet and decuplet baryons, and their radial excitations.
Row~1: Baryon ground-states.
Row~2: Ground-state masses obtained using the ESR described in connection with Eq.\,\eqref{eqESR}.
Row~5: First positive-parity excitations of the ground-states.
Row~6: Masses of positive-parity excitations obtained using the ESR described in connection with Eq.\,\eqref{eqESR}.
Masses in rows labelled ``expt.'' are taken from Ref.\,\protect\cite{Tanabashi:2018oca}; and those in rows labelled ``ESR$_{\rm expt.}$'' were obtained using the ESR discussed in connection with Eq.\,\eqref{eqESR} applied to the empirical masses.  (Underlined entries were used as the basis for associated ESR estimates and the entries marked by asterisks are described in the text surrounding Eqs.\,\eqref{PredictedBWmasses}.)
A hyphen in any position indicates that no empirically known resonance can confidently be associated with the theoretically predicted state.
(All dimensioned quantities are listed in GeV.)
}
\begin{center}
\begin{tabular*}
{\hsize}
{
l@{\extracolsep{0ptplus1fil}}
l@{\extracolsep{0ptplus1fil}}
l@{\extracolsep{0ptplus1fil}}
|l@{\extracolsep{0ptplus1fil}}
l@{\extracolsep{0ptplus1fil}}
l@{\extracolsep{0ptplus1fil}}
l@{\extracolsep{0ptplus1fil}}
|l@{\extracolsep{0ptplus1fil}}
l@{\extracolsep{0ptplus1fil}}
l@{\extracolsep{0ptplus1fil}}
l@{\extracolsep{0ptplus1fil}}}\hline
 & Row & & \rule{0ex}{2.5ex}
$N$ & $\Lambda$ & $\Sigma$ & $\Xi$
   & $\Delta$ & ${\Sigma^\ast}$ & ${\Xi^\ast}$ & $\Omega$ \\\hline
 n=0 & 1 & \rule{0ex}{2.5ex}
DSE & \underline{1.19(13)} & 1.37(14) & 1.41(14) & \underline{1.58(15)} $\ $& \underline{1.35(12)} & 1.52(14) & 1.71(15) &\underline{1.93(17)}\rule{0em}{2.5ex} \\
     & 2 & \rule{0ex}{2.5ex}
 ESR$_{\rm DSE}\ $ &  1.19(13) & 1.39(14) & 1.39(14) & 1.58(15) $\ $
                            & 1.35(12) & 1.54(14) & 1.74(15) & 1.93(17)\rule{0em}{2.5ex} \\
 & 3 & \rule{0ex}{2.5ex}
expt. & \underline{0.94} & 1.12 & 1.19 & \underline{1.31} & \underline{$1.23$} & $1.38$ & $1.53$ & \underline{1.67}\\
& 4 &  \rule{0ex}{2.5ex}
ESR$_{\rm expt.}$ & 0.94 & 1.13 & 1.13 &  1.31 & $1.23$ & $1.39$ & $1.52$ & 1.67\\\hline
 n=1 & 5 & \rule{0ex}{2.5ex}
DSE &  \underline{$1.73(10)$} & $1.85(09)$ & $1.88(11)$ & \underline{$1.99(11)$} &
\underline{$1.79(12)$} & $1.93(11)$ & $2.08(12)$ & \underline{$2.23(13)$}\rule{0em}{2.5ex} \\
 & 6 & \rule{0ex}{2.5ex}
ESR$_{\rm DSE}$ & $1.73(10)$ & $1.86(10)$ & $1.86(10)$ & $1.99(11)$ &
$1.79(12)$ & $1.93(12)$ & $2.08(12)$ & $2.23(13)$\rule{0em}{2.5ex} \\
 & 7 & \rule{0ex}{2.5ex}
expt. &  \underline{$1.44(03)$} & \underline{$1.60^{+0.10}_{-0.04}$} & \underline{$1.66(03)$} &  - & \underline{$1.57(07)$} & \underline{$1.73(03)$} & - & -\\
 & 8 & \rule{0ex}{2.5ex}
ESR$_{\rm expt.}$ &   $1.44(03)$ & $1.64(05)$ & $1.64(05)$ &  $1.84(08)^\ast$
        & $1.57(07)$ & $1.73(03)$ & $1.89(04)^\ast$ & $2.05(02)^\ast$ \\\hline
\end{tabular*}
\end{center}
\end{table*}

\smallskip

\hspace*{-\parindent}(\emph{i})\,{Octet baryons}:\\
{\allowdisplaybreaks
\begin{subequations}
\label{Lidentifications8}
\begin{align}
^2\!S: & \quad {\mathpzc S}^1, {\mathpzc A}^2_\nu, ({\mathpzc A}^3_\nu+{\mathpzc A}^5_\nu) \,;\\
%
^2\!P: & \quad {\mathpzc S}^2, {\mathpzc A}^1_\nu, ({\mathpzc A}^4_\nu+{\mathpzc A}^6_\nu)\,;\\
%
^4\!P: & \quad (2{\mathpzc A}^4_\nu-{\mathpzc A}^6_\nu)/3\,;\\
%
^4\!D:  & \quad  (2{\mathpzc A}^3_\nu-{\mathpzc A}^5_\nu)/3  \,;
\end{align}
\end{subequations}}
\hspace*{-0.5\parindent}\emph{viz}.\ the scalar functions associated with these combinations of Dirac matrices in a Faddeev wave function possess the identified angular momentum correlation between the quark and diquark.
Those functions are:
\begin{subequations}
\label{LFunctionIdentifications}
\begin{align}
\label{LFunctionIdentificationsa}
^2\!S: & \quad \tilde{\mathpzc s}_{\underline{8} 1}^j, \tilde{\mathpzc a}_{\underline{8} 2}^g,
    (\tilde{\mathpzc a}_{\underline{8} 3}^g+2\tilde{\mathpzc a}_{\underline{8} 5}^g)/3\,;\\
^2\!P: & \quad \tilde{\mathpzc s}_{\underline{8} 2}^j, \tilde{\mathpzc a}_{\underline{8} 1}^g,
    (\tilde{\mathpzc a}_{\underline{8} 4}^g+2\tilde{\mathpzc a}_{\underline{8} 6}^g)/3\,;\\
^4\!P: & \quad (\tilde{\mathpzc a}_{\underline{8} 4}^g-\tilde{\mathpzc a}_{\underline{8} 6}^g)\,;\\
^4\!D: & \quad (\tilde{\mathpzc a}_{\underline{8} 3}^g - \tilde{\mathpzc a}_{\underline{8} 5}^g) \,.
\end{align}
\end{subequations}

\hspace*{-\parindent}(\emph{ii})\,{Decuplet baryons}:\\
{\allowdisplaybreaks
\begin{subequations}
\label{Lidentifications10}
\begin{align}
^4\!S: & \quad {\mathpzc D}^1_{\nu\rho}\,;\\
^2\!P: & \quad {\mathpzc D}^4_{\nu\rho}, -({\mathpzc D}^5_{\nu\rho}+{\mathpzc D}^7_{\nu\rho})\,;\\
^4\!P: & \quad {\mathpzc D}^2_{\nu\rho}+(2{\mathpzc D}^5_{\nu\rho}+2{\mathpzc D}^7_{\nu\rho})/3\,;\\
^2\!D:  & \quad  {\mathpzc D}^3_{\nu\rho}, -({\mathpzc D}^6_{\nu\rho}+{\mathpzc D}^8_{\nu\rho})  \,;\\
^4\!D:  & \quad  (-2{\mathpzc D}^6_{\nu\rho}+{\mathpzc D}^8_{\nu\rho}-{\mathpzc D}^1_{\nu\rho})/3  \,;\\
^4\!F:  & \quad  (-4{\mathpzc D}^5_{\nu\rho}+{\mathpzc D}^7_{\nu\rho}-{\mathpzc D}^2_{\nu\rho})/5  \,.
\end{align}
\end{subequations}}
The associated scalar functions are
{\allowdisplaybreaks
\begin{subequations}
\label{LFunctionIdentifications10}
\begin{align}
^4\!S: & \quad \tilde{\mathpzc a}_{\underline{10} 1}^g - (\tilde{\mathpzc a}_{\underline{10} 6}^g-\tilde{\mathpzc a}_{\underline{10} 8}^g)/3\,;\\
^2\!P: & \quad \tilde{\mathpzc a}_{\underline{10} 4}^g, (2\tilde{\mathpzc a}_{\underline{10} 2}^g-\tilde{\mathpzc a}_{\underline{10} 5}^g-2\tilde{\mathpzc a}_{\underline{10} 7}^g)/3\,;\\
^4\!P: & \quad
\tilde{\mathpzc a}_{\underline{10} 2}^g-(\tilde{\mathpzc a}_{\underline{10} 5}^g-\tilde{\mathpzc a}_{\underline{10} 7}^g)/5\,;\\
^2\!D:  & \quad  \tilde{\mathpzc a}_{\underline{10} 3}^g,
 -(\tilde{\mathpzc a}_{\underline{10} 6}^g+2\tilde{\mathpzc a}_{\underline{10} 8}^g)/3\,;\\
^4\!D:  & \quad  -\tilde{\mathpzc a}_{\underline{10} 6}^g+\tilde{\mathpzc a}_{\underline{10} 8}^g  \,;\\
^4\!F:  & \quad  -\tilde{\mathpzc a}_{\underline{10} 5}^g+\tilde{\mathpzc a}_{\underline{10} 7}^g  \,.
\end{align}
\end{subequations}}

\section{Solutions and their Properties}
\label{solutions}
\subsection{Masses of the dressed-quark cores}
\label{SecSpectrum}
We can now report results obtained by solving the Faddeev equations for octet and decuplet baryons and their first positive-parity excitations.
Our computed masses are listed in Table~\ref{OctetDecupletMasses}: the uncertainties indicate the response of a given mass to a coherent 5\% increase/decrease in the mass-scales associated with the diquarks and dressed-quarks, Eqs.\,\eqref{dqmasses}, \eqref{lambdaval}, respectively.

It is worth noting the emergence of a $\Sigma$-$\Lambda$ mass-splitting despite the fact that we have assumed isospin symmetry, \emph{i.e}.\ mass-degenerate $u$- and $d$-quarks, described by the same propagator, so that all diquarks in an isospin multiplet are degenerate.  The origin of the $\Sigma$-$\Lambda$ splitting can be understood by comparing Eqs.\,\eqref{flavourLambda} and \eqref{flavourSigma}, with the latter adapted to the neutral $\Sigma^0$ case following Eq.\,(49) in Ref.\,\cite{Chen:2012qr}.  Whilst the $\Lambda^0$ and $\Sigma^0$ baryons are associated with the same combination of valence-quarks,
their spin-flavour wave functions are different: the $\Lambda_{I=0}^0$ contains more of the lighter $J=0$ diquark correlations than the $\Sigma_{I=1}^0$.  It follows that the $\Lambda^0$ must be lighter than the $\Sigma^0$.
The mechanism underlying this splitting is analogous to that which produces the $\pi$-$\rho$ mass difference, and also to that associated with the colour-hyperfine interaction used in quark models.  It is realised here via the breaking of isospin-symmetry in the associated baryon wave functions.

\begin{figure}[t]
\includegraphics[clip, width=0.45\textwidth]{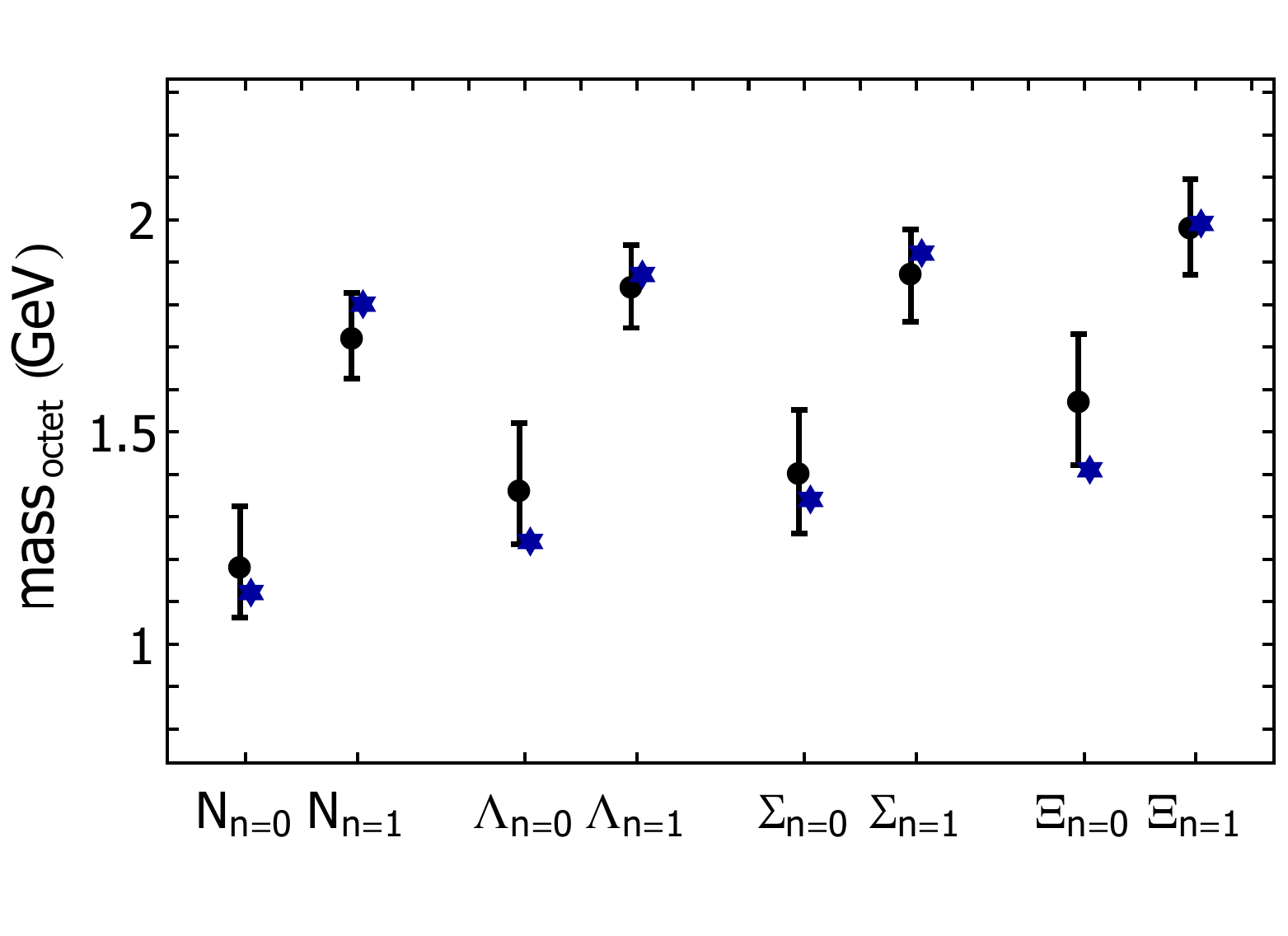}
\includegraphics[clip, width=0.45\textwidth]{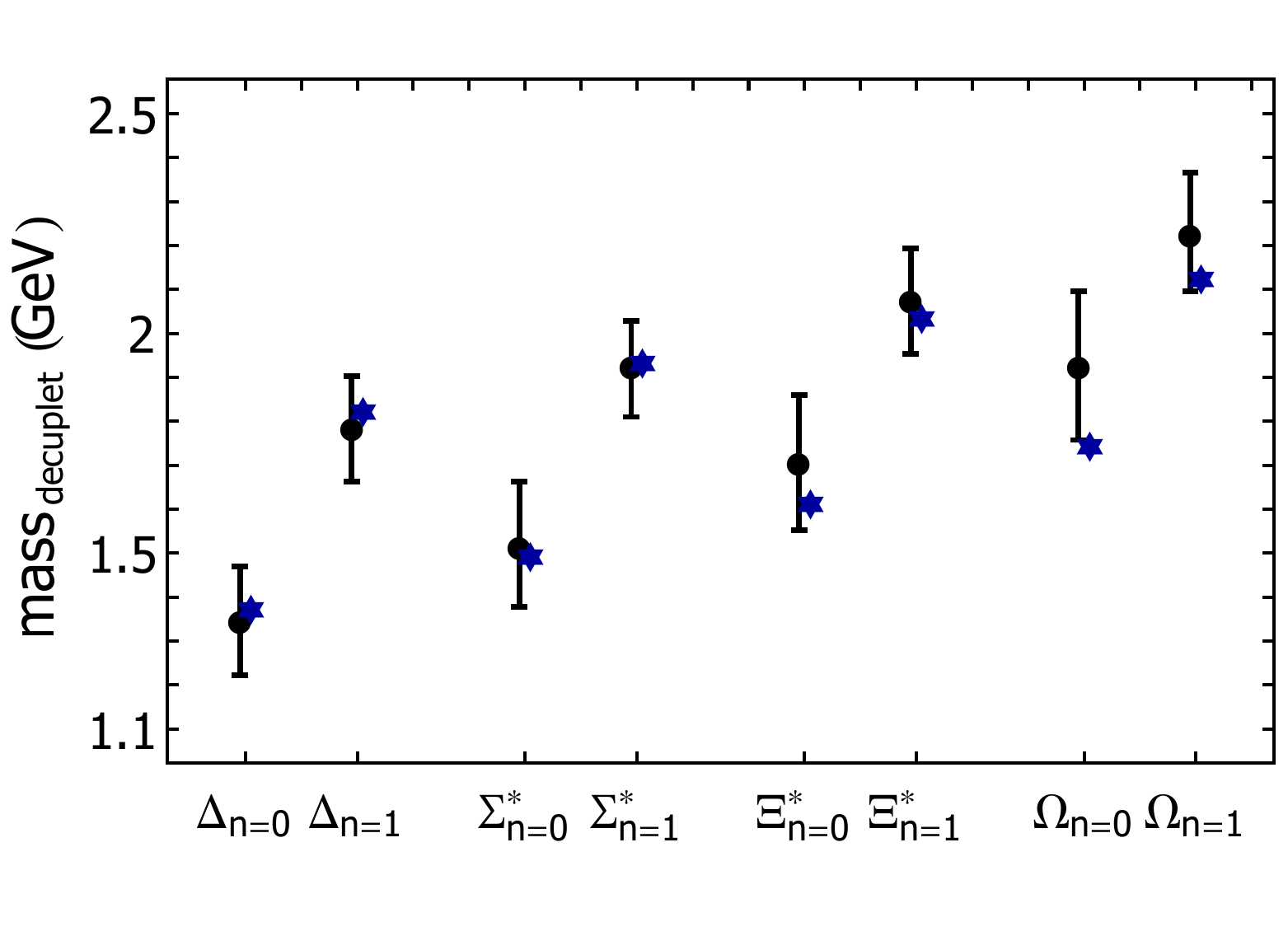}
\caption{\label{CompareYa}
Comparison between the masses listed in Table~\ref{OctetDecupletMasses} (black circles), computed herein using Faddeev equation kernels built with dressed-quarks and diquarks described by QCD-like momentum-dependent propagators and amplitudes, and those obtained using a symmetry-preserving treatment of a vector$\,\times\,$vector contact-interaction (blue stars) \cite{Lu:2017cln}.
\textbf{Upper panel}: octet states.
\textbf{Lower panel}: decuplet states.
The vertical riser indicates the response of our results to a coherent $\pm 5$\% change in the mass-scales associated with the diquarks and dressed-quarks, Eqs.\,\eqref{dqmasses}, \eqref{lambdaval}, respectively.
The horizontal axis lists a particle name with a subscript that indicates whether it is ground-state ($n=0$) or first positive-parity excitation ($n=1$).
}
\end{figure}

\begin{figure}[t]
\includegraphics[clip, width=0.45\textwidth]{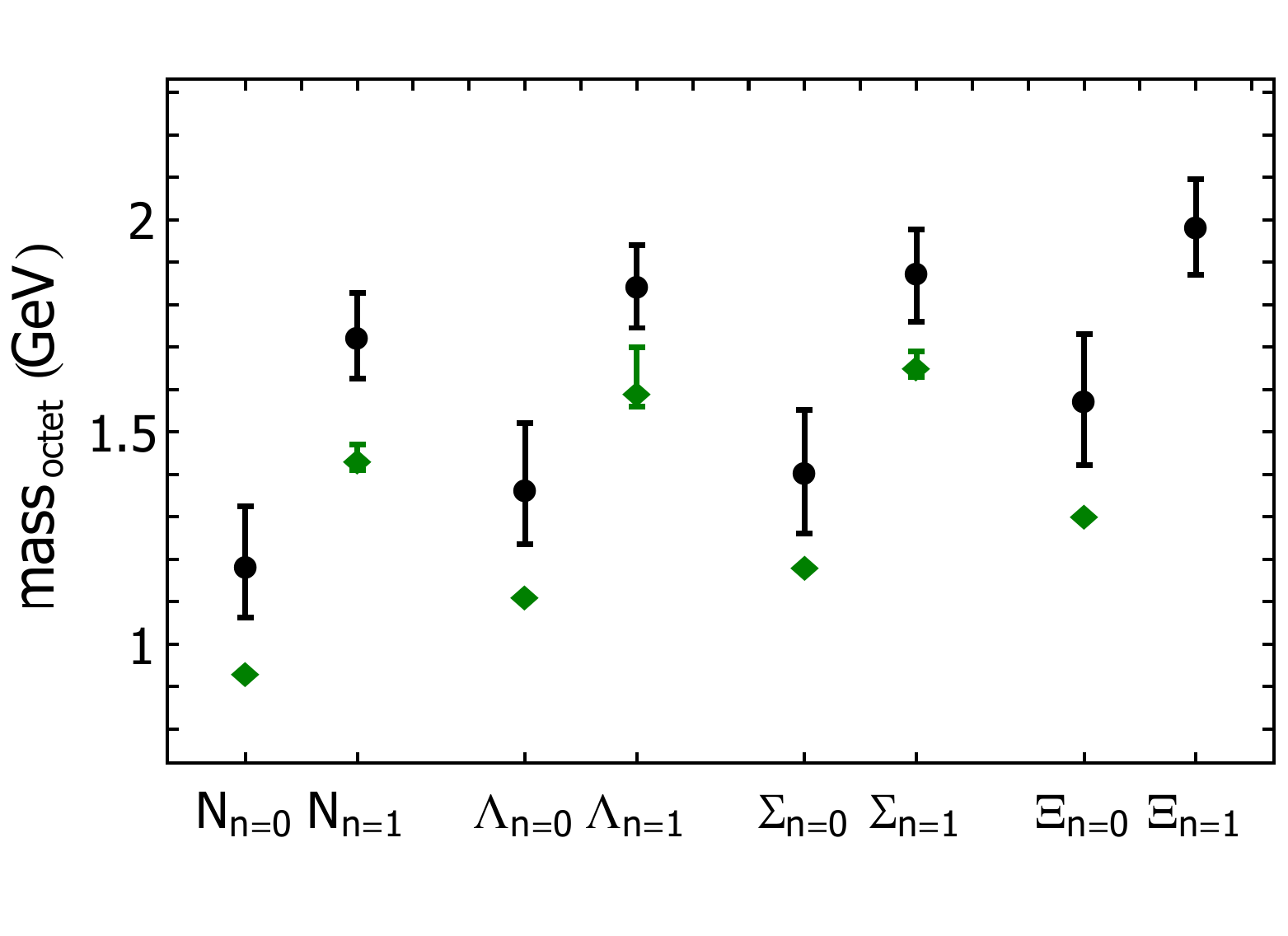}
\includegraphics[clip, width=0.45\textwidth]{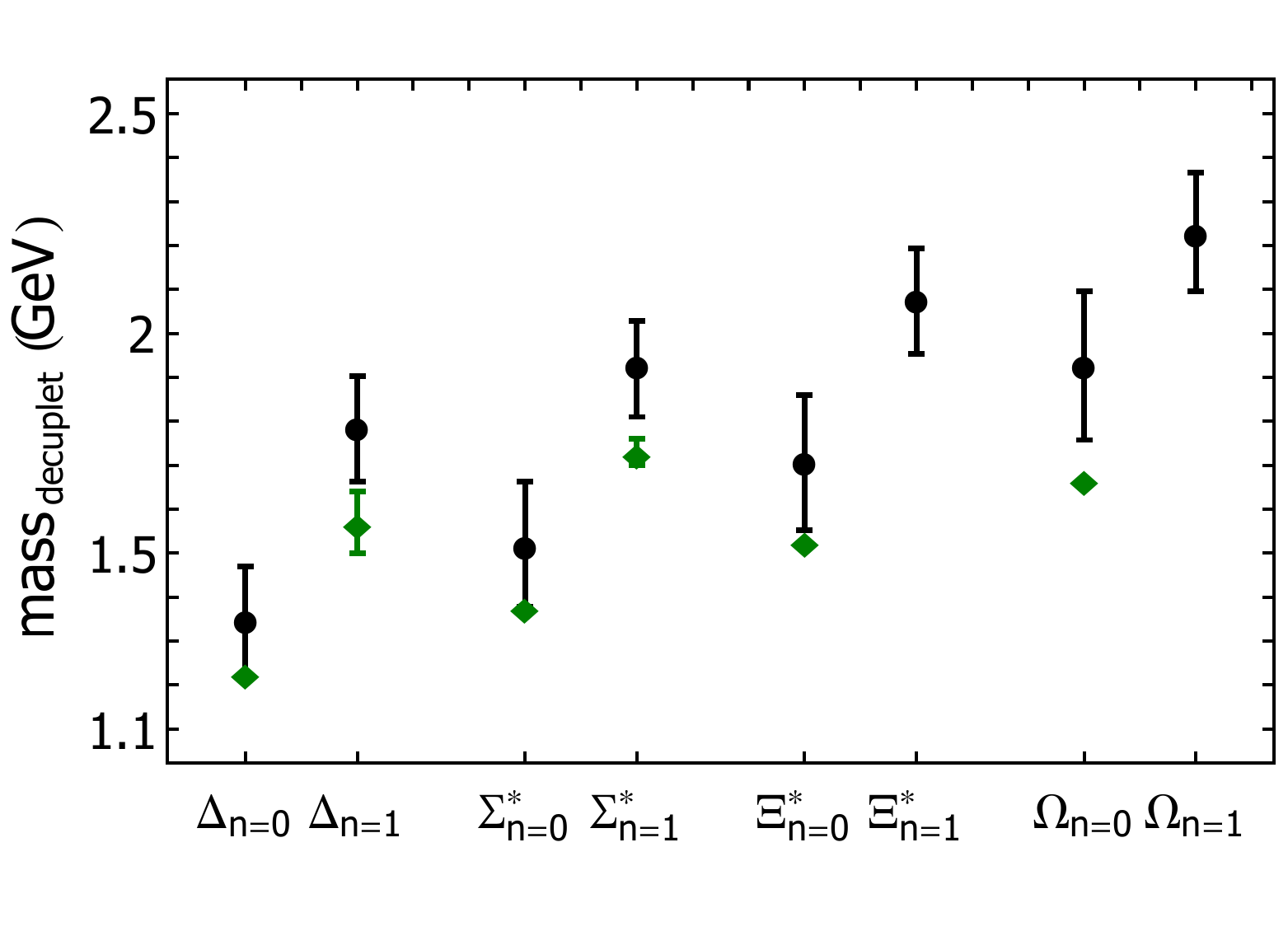}
\caption{\label{BarLD}
\textbf{Upper panel}: Pictorial representation of octet masses in Table~\protect\ref{OctetDecupletMasses}.  \emph{Circles} (black) -- computed masses.
The vertical riser indicates the response of our predictions to a coherent $\pm 5$\% change in the mass-scales associated with the diquarks and dressed-quarks, Eqs.\,\eqref{dqmasses}, \eqref{lambdaval}, respectively.
\emph{Diamonds} (green) -- empirical Breit-Wigner masses \cite{Tanabashi:2018oca}.
The horizontal axis lists a particle name with a subscript that indicates whether it is ground-state ($n=0$) or first positive-parity excitation ($n=1$).
\textbf{Lower panel}: Analogous plot for the decuplet masses in Table~\protect\ref{OctetDecupletMasses}.
Where noticeable, the estimated uncertainty in the location of a resonance's Breit-Wigner mass is indicated by an error bar.}
\end{figure}

The quark-core masses of octet and decuplet baryons and their first positive-parity excitations have previously been computed using the SPCI \cite{Lu:2017cln}.  Thus, in Fig.\,\ref{CompareYa}, we compare our predictions with those reported therein.  Evidently, so far as computed masses are concerned, there is little material difference between the results obtained using Faddeev equation kernels with realistic momentum dependence and those produced by a momentum-independent interaction.  This outcome is consistent with findings elsewhere \cite{GutierrezGuerrero:2010md, Roberts:2010rn, Roberts:2011wy, Wilson:2011aa, Chen:2012txa, Segovia:2013rca, Segovia:2014aza, Xu:2015kta, Bedolla:2015mpa, Bedolla:2016yxq, Serna:2017nlr, Raya:2017ggu}, \emph{viz}.\ implemented judiciously, the SPCI typically produces results that are practically equivalent to those obtained with QCD-like kernels so long as the probe momenta involved are smaller than the relevant dressed-quark masses, Eqs.\,\eqref{MEq}.

\begin{figure*}[!t]
\centerline{%
\includegraphics[clip, width=0.91\textwidth]{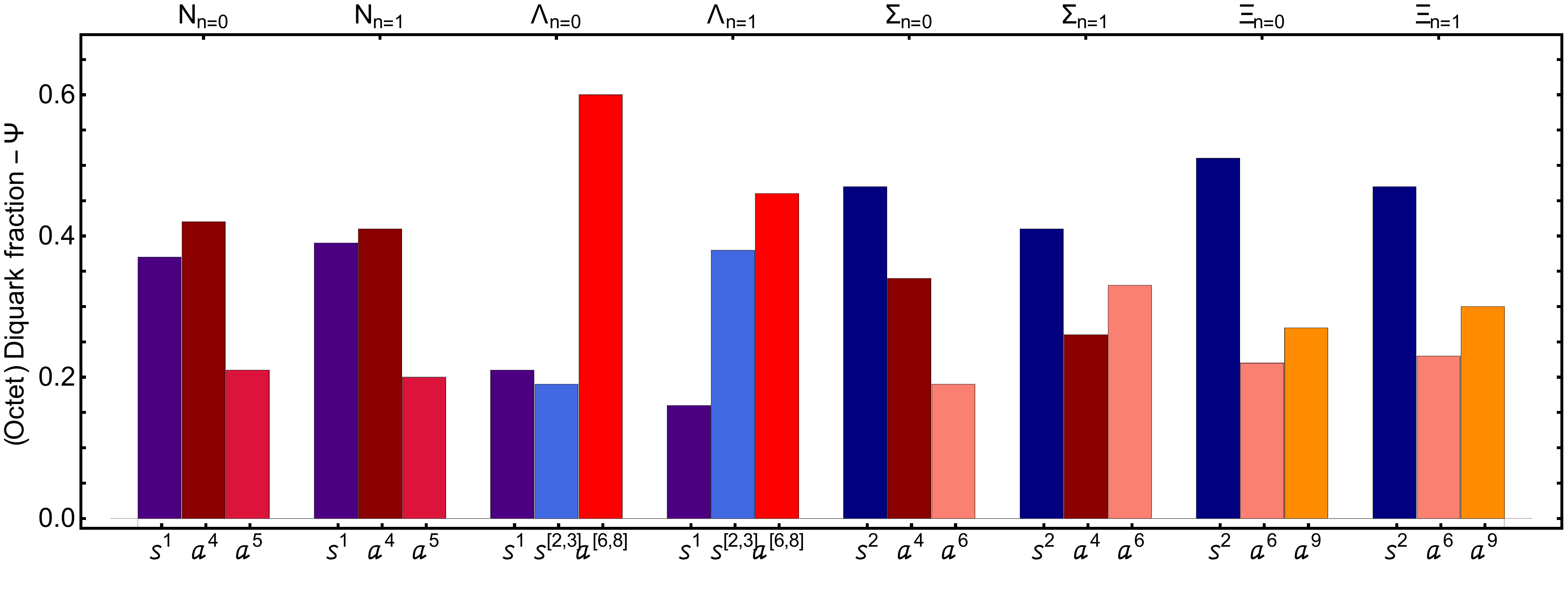}}
\centerline{%
\includegraphics[clip, width=0.91\textwidth]{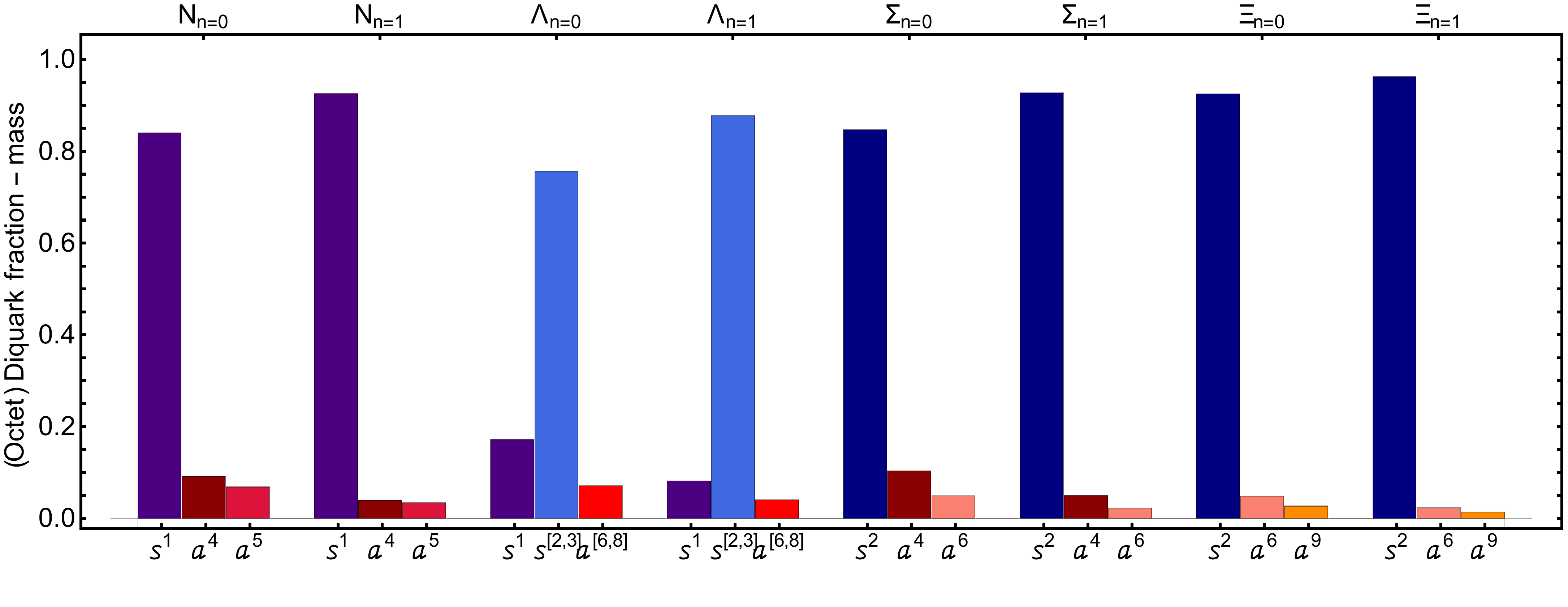}}
\caption{\label{BarLDN}
\emph{Upper panel} -- Relative strengths of various diquark components within the indicated baryon's Faddeev amplitude, defined via Eqs.\,\eqref{mathbbD}, \eqref{Dfracs}.
%
\emph{Lower panel} -- Relative contribution to a baryon's mass from a given diquark correlation in that baryon's Faddeev amplitude, defined in association with Eqs.\,\eqref{qqPmasses}.
}
\end{figure*}

Table~\ref{OctetDecupletMasses} compares our predictions for the quark-core masses with empirical values, where known, of the Breit-Wigner mass associated with each state considered \cite{Tanabashi:2018oca}.\footnote{
Table~\ref{OctetDecupletMasses} associates the nucleon's first positive-parity excitation with the dressed-quark core of the Roper resonance.  The justification for this identification is presented in Ref.\,\cite{Burkert:2017djo}, along with a discussion of the attendant controversy.}
For added clarity, this information is also depicted in Fig.\,\ref{BarLD}.  It is apparent that the computed masses are uniformly larger than the corresponding empirical values.  As described in Sec.\ref{FEremarks}, this is because our results should be viewed as those of a given baryon's dressed-quark core, whereas the empirical values include all contributions, including MB\,FSIs, which typically generate a measurable reduction \cite{Suzuki:2009nj}.  This was explained and illustrated in a study of the nucleon, its parity-partner and their radial excitations in Ref.\,\cite{Chen:2017pse}; and has also been demonstrated using the SPCI \cite{Chen:2012qr, Lu:2017cln}.  Identifying the difference between our predictions and experiment as the result of MB\,FSIs, then one finds that such effects are fairly homogeneous across the spectrum:
\begin{subequations}
\begin{align}
{\rm mean}[m_{\rm DSE}^{\underline 8} - m_{\rm expt.}^{\underline 8}] &= 0.26(4)\,, \\
{\rm mean}[m_{\rm DSE}^{\underline{10}} - m_{\rm expt.}^{\underline{10}}] & = 0.19(5)\,.
\end{align}
\end{subequations}
Namely, they act to reduce the mass of ground-state octet and decuplet baryons and their first positive-parity excitations by $0.23(6)\,$GeV.

Here it is also worth recalling an equal spacing rule (ESR) \cite{Okubo:1961jc, GellMann:1962xb}; namely, within a given level, to a good approximation, baryon masses can be related linearly to number-weighted combinations of mass-scales associated with their constituent valence-quarks \cite{Qin:2018dqp}.  This is illustrated in the second row of Table~\ref{OctetDecupletMasses}, which were obtained by defining
\begin{subequations}
\label{eqESR}
\begin{eqnarray}
M_u^{\underline{8} 0}  & = m_N/3 & = 0.397(43)\,,\; \\
M_s^{\underline{8} 0}  & = (m_\Xi-M_u^{\underline{8} 0})/2 \; & = 0.592(53)\,,\\
M_u^{\underline{10} 0} &= m_\Delta/3 & = 0.450(40)\,,\;\\
M_s^{\underline{10} 0}  & = m_\Omega/3 & = 0.643(57)\,,
\end{eqnarray}
\end{subequations}
and identifying, \emph{e.g}.\ $m_{\Sigma^{\ast}}^{\rm ESR} = 2 M_u^{\underline{10} 0}+M_s^{\underline{10} 0}$.  Repeating such a procedure with the empirical entries in Row~3, we obtain the fourth row in Table~\ref{OctetDecupletMasses}.  Evidently, ESR estimates are reliable.

\begin{figure*}[t!]
\centerline{%
\includegraphics[clip, width=0.9\textwidth]{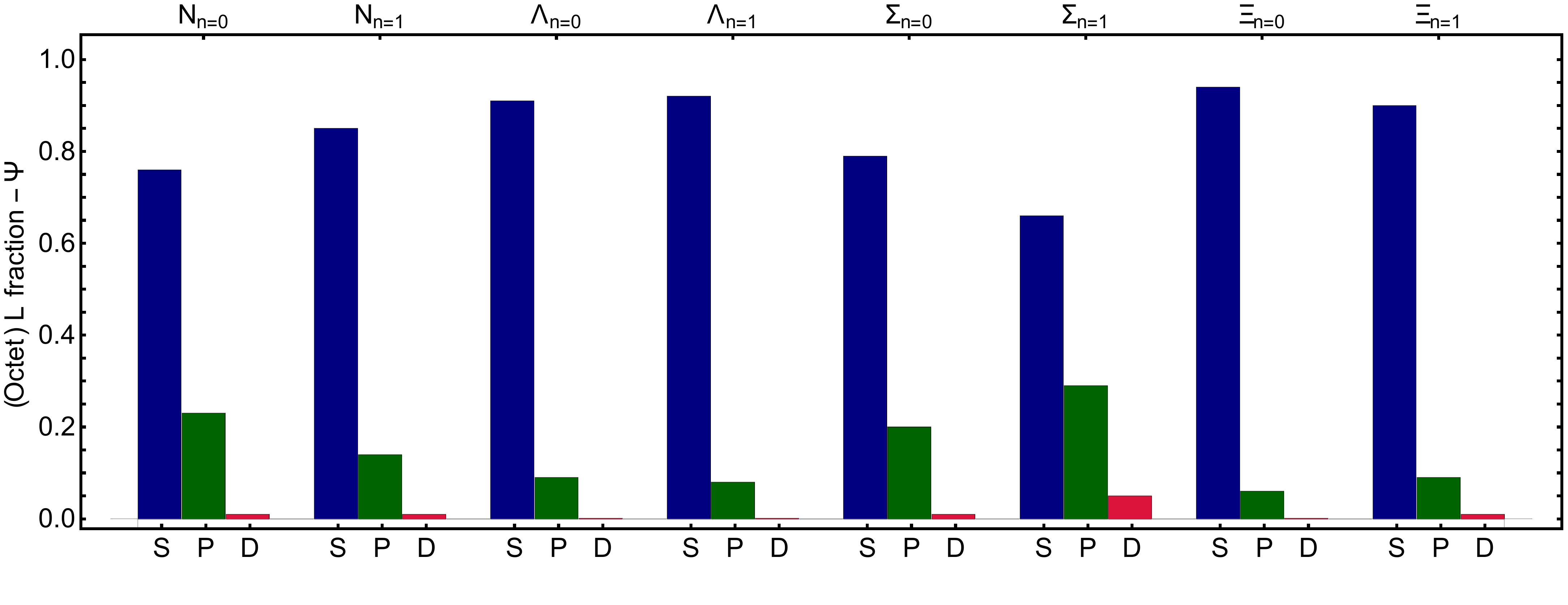}}
\centerline{%
\includegraphics[clip, width=0.9\textwidth]{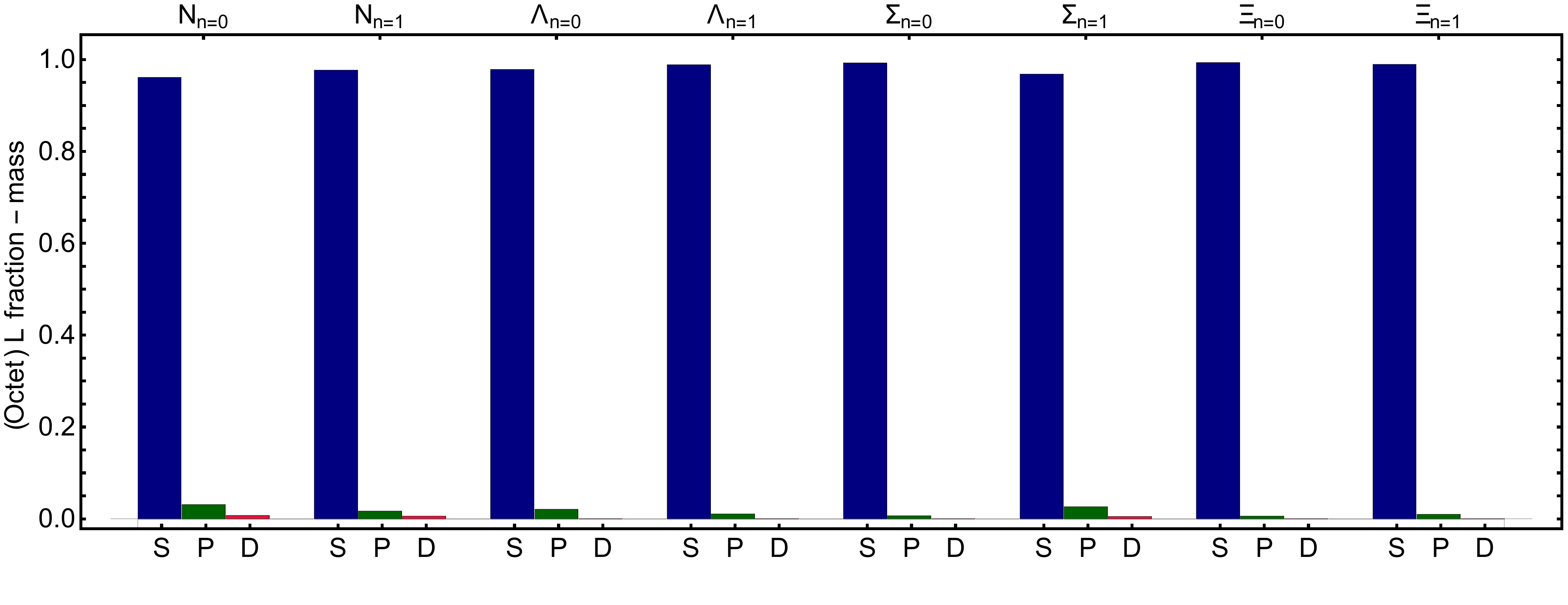}}
\caption{\label{Loctet}
Octet baryons and their first positive-parity excitations.
\textbf{Upper panel} -- Baryon rest-frame quark-di\-quark orbital angular momentum fractions, as defined in Eqs.\,\eqref{Lfracs}.
\textbf{Lower panel} -- Relative contribution of various quark-diquark orbital angular momentum components to the mass of a given baryon.
}
\end{figure*}

We also employed the ESR in connection with our predictions for the quark-core masses of the baryons' first positive-parity excitations, with the results in Rows~5 and 6 of Table~\ref{OctetDecupletMasses}.  Once again, apart from missing the isospin-breaking $\Sigma_{n=1}-\Lambda_{n=1}$ splitting, all other masses are accurately reproduced.  (Even the $\Lambda_{n=1}$ and $\Sigma_{n=1}$ masses are separately accurate to within 1\%.)
These outcomes suggest that reasonable estimates of the empirical masses for the first positive-parity excitations of all octet and decuplet baryons may be obtained by using ESRs based on measured masses.
Using a $(\Lambda,\Sigma)$-isospin average to define the empirical ESR for octet radial excitations, this approach yields the results marked by asterisks in Row~8 of Table~\ref{OctetDecupletMasses}, \emph{i.e}.\ our Faddeev equation analysis predicts the existence of positive-parity excitations of the $\Xi$, $\Xi^\ast$, $\Omega$ baryons with the following Breit-Wigner masses (in GeV):
\begin{subequations}
\label{PredictedBWmasses}
\begin{align}
m_{\Xi_{n=1}} & = 1.84(08)\,,\\
m_{\Xi_{n=1}^\ast} & = 1.89(04)\,,\;
m_{\Omega_{n=1}} = 2.05(02)\,.
\end{align}
\end{subequations}
Such states are also found in quark models, \emph{e.g}.\  Refs.\,\cite{Capstick:1986bm, Valcarce:2005rr, Santopinto:2014opa, Yang:2017qan}, but the mass values are model-dependent.

Since these states possess strangeness $-2$ ($\Xi$, $\Xi^\ast$) and $-3$ ($\Omega$), finding them in photoproduction and electron scattering experiments is difficult owing to marked Zweig-rule suppression and the need to measure multi-particle ($>3$) final states.  Notwithstanding that, a search for very strange baryons is part of the programmes with the CLAS12 and GlueX detectors at Jefferson Lab (JLab) \cite{E12-11-005A, E12-13-003, Amaryan:2017ldw}.
They may also be located using kaon beams at the Japan proton accelerator research complex (J-PARC) \cite{Kamano:2015hxa, Briscoe:2015qia}.
It is possible that signals for the predicted $\Xi_{n=1}$ state may already be present in data obtained with $\mu^+$-beams using the COMPASS detector at CERN \cite{Adolph:2013dhv} and all predictions in Eq.\,\eqref{PredictedBWmasses} could be tested at CERN in future using the proposed kaon-enriched hadron beams \cite{Denisov:2018unj}.

\subsection{Diquark content}
\label{SecDiquark}
It is interesting now to dissect the results in various ways and thereby sketch the character of the quark cores that constitute the ground-state octet and decuplet baryons and their first positive-parity excitations.  We begin by exposing their diquark content.

\begin{figure*}[t!]
\centerline{%
\includegraphics[clip, width=0.9\textwidth]{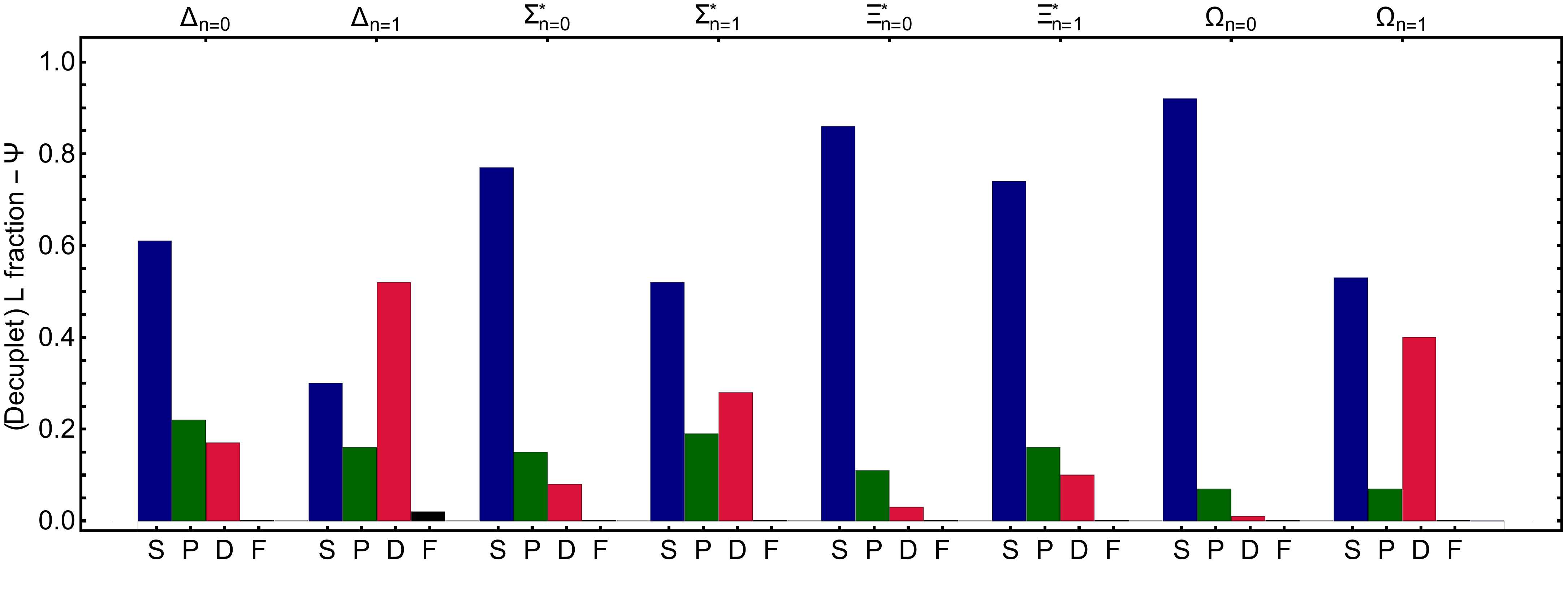}}
\centerline{%
\includegraphics[clip, width=0.9\textwidth]{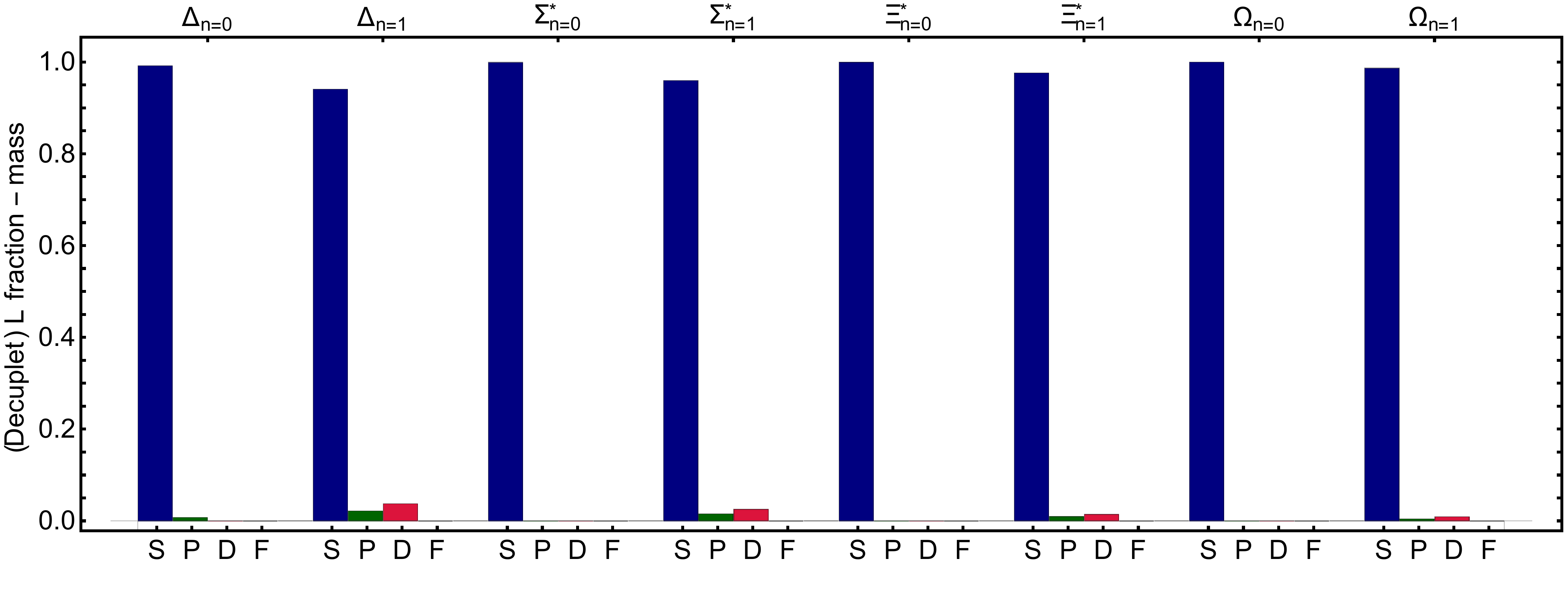}}
\caption{\label{Ldecuplet}
Decuplet baryons and their first positive-parity excitations.
\textbf{Upper panel} -- Baryon rest-frame quark-di\-quark orbital angular momentum fractions, as defined in Eqs.\,\eqref{Lfracs}.
\textbf{Lower panel} -- Relative contribution of various quark-diquark orbital angular momentum components to the mass of a given baryon.
}
\end{figure*}

The Faddeev amplitude of each baryon can be decomposed into a sum of $N_{qq}$ terms, $\{{\mathpzc F}_i,i=1,\ldots,N_{qq}\}$, each one of which is directly identifiable with a particular diquark component.  The value of $N_{qq}$ depends on the baryon's spin-flavour structure, Eqs.\,\eqref{FaddeevSpinFlavour} and \,\eqref{DecupletSpinFlavour}:
\begin{equation}
\rule{-0.9em}{0ex}
\begin{array}{l|cccc|cccc}
    & N & \Lambda & \Sigma & \Xi  & \Delta & \Sigma^\ast & \Xi^\ast & \Omega
    \\\hline
N_{qq} & 14 & 10 & 14 & 14 & 8 & 16 & 16 & 8
\\
\end{array}.
\end{equation}
%
In connection with each term, we define
\begin{equation}
\label{mathbbD}
{\mathbb D}_i = \int\frac{d^4\ell}{(2\pi)^4}\,|{\mathpzc F}_i(\ell^2,\ell\cdot P)|^2
\end{equation}
and subsequently compute
\begin{equation}
\label{Dfracs}
{\mathbb Q}_{\mathpzc t}  ={\mathbb W}^{-1} \,  \sum_{i \in {\mathpzc t}} {\mathbb D}_i \,,\;
{\mathbb W}  = \sum_{i=1}^{N_{qq}}\, {\mathbb D}_i\,,
\end{equation}
where ${\mathpzc t}$ ranges over the ${\mathpzc s}(=0^+)$ and ${\mathpzc a}(=1^+)$ components of the baryon considered.  Here, ${\mathbb W}$ defines a four-dimensional $L^2$-norm of the baryon's Faddeev amplitude and the ratios ${\mathbb Q}_{t={\mathpzc s}^j,{\mathpzc a}^g}$ express the relative size of the contribution from each diquark correlation to this norm.  The values of these fractions are one indication of the relative strengths of the various diquark components within a baryon.  They are listed in Table~\ref{FaddeevAmps} and depicted for the more complex octet states in Fig.\,\ref{BarLDN}--upper panel.

An alternative gauge is to consider the relative contribution to a given baryon's mass owing to each of the diquark components in its Faddeev amplitude.  We evaluate this by computing the hadron's mass in the absence of all except the dominant diquark correlation (often the lightest possible contributor) and then introducing the remaining correlations in their order of importance, which is determined by trial-and-error.
Typically, the mass obtained using only the dominant correlation is larger than the all-correlation result.
(Otherwise, the Faddeev equation would suppress any new correlation.)
We therefore define the relative mass-contribution of a given correlation as follows.
Suppose two diquarks contribute: $qq_1$, $qq_2$, with $qq_1$ dominant.
Further suppose that the $qq_1$-only baryon has mass $m_1$ and adding $qq_2$ gives $m_{\rm final} = m_2<m_1$, then
\begin{subequations}
\label{qqPmasses}
\begin{align}
P_{1} & = m_1/(m_1+|m_2-m_1|)\,,\; \\
P_{2} & = |m_2-m_1|/(m_1+|m_2-m_1|)\,.
\end{align}
\end{subequations}
This procedure has a clear generalisation to systems with more than two diquark correlations; and the results obtained in this way are listed in Table~\ref{qqmasses} and depicted for the more complicated octet states in Fig.\,\ref{BarLDN}--lower panel.

As observed elsewhere \cite{Chen:2017pse}, the difference between the upper and lower panels of Fig.\,\ref{BarLDN} is marked.  In each of the octet cases depicted in the lower panel, there is a single dominant diquark component; namely, a scalar diquark; and each new correlation adds binding, reducing the computed mass.
On the other hand, measuring the relative strength of diquark correlations using the Faddeev amplitude decomposition, drawn in Fig.\,\ref{BarLDN}--upper panel, one arrives at a somewhat different picture; but such differences can largely be attributed to the lack of interference between diquark components in the measure defined by Eqs.\,\eqref{mathbbD}, \eqref{Dfracs}.
Notwithstanding these facts, comparisons between baryons using any single measure are meaningful, \emph{e.g}.\ using either scheme, the nucleon and its first positive-parity excitation possess very similar diquark content.

Naturally, the true importance of a given correlation within one or another baryon is properly measured by its role in determining observables, and this speaks in favour of the mass-fraction measure and its analogues, such as the canonical normalisation \cite{Segovia:2015hra}, which relates to electric charge contributions.
Following this reasoning, impacts of various diquark components of the nucleon and $\Delta$-baryon on their elastic and transition form factors have been explored elsewhere \cite{Segovia:2014aza, Segovia:2015hra, Segovia:2015ufa, Segovia:2016zyc, Chen:2018nsg}.

\subsection{Rest-frame orbital angular momentum}
\label{SecJLS}
Drawing upon Sec.\,\ref{SecFWFs}, we now expose the rest-frame orbital angular momentum content of each baryon.  Connected with each matrix in Eqs.\,\eqref{Lidentifications8}, \eqref{Lidentifications10}, there is a scalar function, the collection of which we denote as $\{{\mathpzc Y}_i,i=1,\ldots,N_{qq}\}$,
\emph{e.g}.\ the five rest-frame $^2\!S$-components in the proton are identified with ${\mathpzc Y}_{1,\ldots,5}$ and, using Eq.\,\eqref{LFunctionIdentificationsa}, these functions are
\begin{equation}
\tilde{\mathpzc s}_{N 1}^1,\;
\tilde{\mathpzc a}_{N 2}^{4,5},\;
[\tilde{\mathpzc a}_{N 3}^{4,5} +2\tilde{\mathpzc a}_{N 5}^{4,5}]/3.
\end{equation}
Using this decomposition, we compiled Table~\ref{tableL8}.
Plainly every one of the systems considered is primarily $S$-wave in nature, since they are not generated by the Faddeev equation unless $S$-wave components are contained in the wave function.
This observation provides support in quantum field theory for the constituent-quark model classifications of these systems, so long as here angular momentum is understood at the hadronic scale to be that between the quark and diquark.
Notwithstanding that, Table~\ref{tableL8} reveals that $P$-wave components play a measurable role in octet ground-states and their first positive-parity excitations: they are attractive in ground-states and repulsive in the excitations.
Regarding decuplet systems: the ground-state masses are almost completely insensitive to non-$S$-wave components; and in the first positive-parity excitations, $P$-wave components generate a little repulsion, some attraction is provided by $D$-waves, and $F$-waves have no measurable impact.

\begin{table}[t]
\caption{\label{tableL8}
Computed baryon quark-core masses: \textbf{upper panel}, flavour-octet; and \textbf{lower panel}, flavour-decuplet.
Row\,1, each panel: results obtained using the complete Faddeev wave function, \emph{i.e}.\ with all angular momentum components included.
Subsequent rows: masses obtained when the indicated rest-frame angular momentum component(s) is(are) excluded from the Faddeev wave function.
Empty locations indicate that a solution is not obtained under the conditions indicated.
(All dimensioned quantities are listed in GeV.)}
\begin{center}
\begin{tabular*}
{\hsize}
{
c@{\extracolsep{0ptplus1fil}}
|c@{\extracolsep{0ptplus1fil}}
c@{\extracolsep{0ptplus1fil}}
c@{\extracolsep{0ptplus1fil}}
c@{\extracolsep{0ptplus1fil}}
c@{\extracolsep{0ptplus1fil}}
c@{\extracolsep{0ptplus1fil}}
c@{\extracolsep{0ptplus1fil}}
c@{\extracolsep{0ptplus1fil}}}\hline
$L\,$content & $N_{n=0}$ & $N_{n=1}$ & $\Lambda_{n=0}$ & $\Lambda_{n=1}$ & $\Sigma_{n=0}$ & $\Sigma_{n=1}$ & $\Xi_{n=0}$ & $\Xi_{n=1}$ \\\hline
  $S,P,D$ & 1.19 & 1.73 & 1.37 & 1.85 & 1.41 & 1.88 & 1.58 & 1.99   \\\hline
 $-,P, D$ & $-$ & $-$ & $-$ & $-$ & $-$ & $-$ & $-$ & $-$ \\
 $S,-, D$ & 1.24 & 1.71 & 1.40 & 1.83 & 1.42 & 1.84 & 1.59 & 1.97 \\
$S, P, -$ & 1.20 & 1.74 & 1.37 & 1.85 & 1.41 & 1.89 & 1.58 & 1.99  \\\hline
 $S,-, -$ & 1.24 & 1.71 & 1.40 & 1.83 & 1.42 & 1.84 & 1.59 & 1.97 \\\hline
\end{tabular*}
\end{center}
\begin{center}
\begin{tabular*}
{\hsize}
{
c@{\extracolsep{0ptplus1fil}}
|c@{\extracolsep{0ptplus1fil}}
c@{\extracolsep{0ptplus1fil}}
c@{\extracolsep{0ptplus1fil}}
c@{\extracolsep{0ptplus1fil}}
c@{\extracolsep{0ptplus1fil}}
c@{\extracolsep{0ptplus1fil}}
c@{\extracolsep{0ptplus1fil}}
c@{\extracolsep{0ptplus1fil}}}\hline
$L\,$content & $\Delta_{n=0}$ & $\Delta_{n=1}$ & $\Sigma^\ast_{n=0}$ & $\Sigma^\ast_{n=1}$ & $\Xi^\ast_{n=0}$ & $\Xi^\ast_{n=1}$ & $\Omega_{n=0}$ & $\Omega_{n=1}$ \\\hline
  $S,P,D,F$ & 1.35 & 1.79 & 1.52 & 1.93 & 1.71 & 2.08 & 1.93 & 2.23   \\\hline
$-, P, D, F$ & $-$ & $-$ & $-$ & $-$ & $-$ & $-$ & $-$ & $-$ \\
$S, -, D, F$ & 1.36 & 1.75 & 1.52 & 1.90 & 1.71 & 2.06 & 1.93 & 2.22 \\
$S, P, -, F$ & 1.35 & 1.82 & 1.52 & 1.95 & 1.71 & 2.09 & 1.93 & 2.24 \\
$S, P, D, -$ & 1.35 & 1.79 & 1.52 & 1.93 & 1.71 & 2.08 & 1.93 & 2.23  \\\hline
$S, -, -, -$ & 1.35 & 1.80 & 1.52 & 1.93 & 1.71 & 2.08 & 1.93 & 2.23  \\\hline
\end{tabular*}
\end{center}
\end{table}

In order to further elucidate these remarks, we turn our attention to the Faddeev wave functions themselves and, for each baryon, compute
\begin{equation}
{\mathbb L}_i = \int \frac{d^4 \ell}{(2\pi)^4} \, |  {\mathpzc Y}_{i}(\ell^2,\ell\cdot P)|^2,
\end{equation}
and subsequently define the following rest-frame angular momentum strengths:
\begin{subequations}
\label{Lfracs}
\begin{align}
&
\begin{array}{ll}
\displaystyle
{\mathbb S} ={\mathbb T}^{-1} \, \sum_{i \in \, S} {\mathbb L}_i \,,
& \displaystyle {\mathbb P} ={\mathbb T}^{-1} \, \sum_{i \in \, P} {\mathbb L}_i \,,\\
\displaystyle
{\mathbb D}  ={\mathbb T}^{-1} \, \sum_{i \in \, D} {\mathbb L}_i \,,
&
\displaystyle
{\mathbb F} ={\mathbb T}^{-1} \, \sum_{i \in \, F} {\mathbb L}_i \,,\\
\end{array}\\
& {\mathbb T}  = \sum_{i=1}^{N_{qq}}\, {\mathbb L}_i\,.
\end{align}
\end{subequations}
Constructed thus, ${\mathbb T}$ defines a four-dimensional $L^2$-norm of the baryon's rest-frame Faddeev wave function and the ratios ${\mathbb S}$, ${\mathbb P}$, ${\mathbb D}$, ${\mathbb F}$ express the relative size of the contribution from each angular momentum component to this norm.  Our results are depicted in the upper panels of Figs.\,\ref{Loctet}, \ref{Ldecuplet}.

As with our examination of the diquark content, another gauge of the relative importance of different partial waves within a baryon is to depict their contributions to a baryon's mass, which can be computed using the information in Table~\ref{tableL8}.  The results are depicted in the lower panels of Figs.\,\ref{Loctet}, \ref{Ldecuplet}.  Once again, even though some contributions are repulsive and others attractive, we draw all bars as positive, following a procedure analogous to that described in connection with Eqs.\,\eqref{qqPmasses}.

\begin{figure*}[!th]
\begin{center}
\begin{tabular}{lr}
\includegraphics[clip,width=0.42\linewidth]{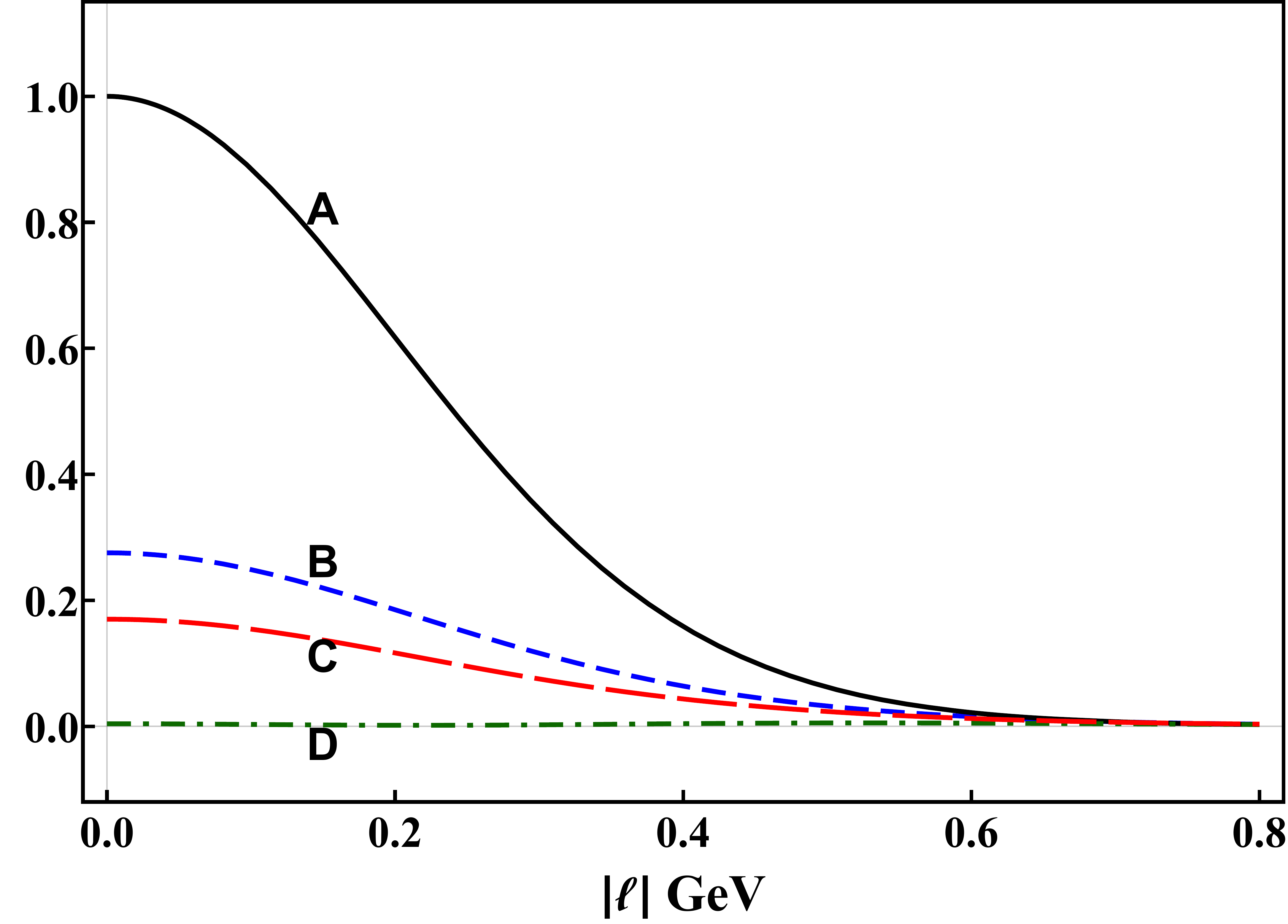}\hspace*{2ex } &
\includegraphics[clip,width=0.43\linewidth]{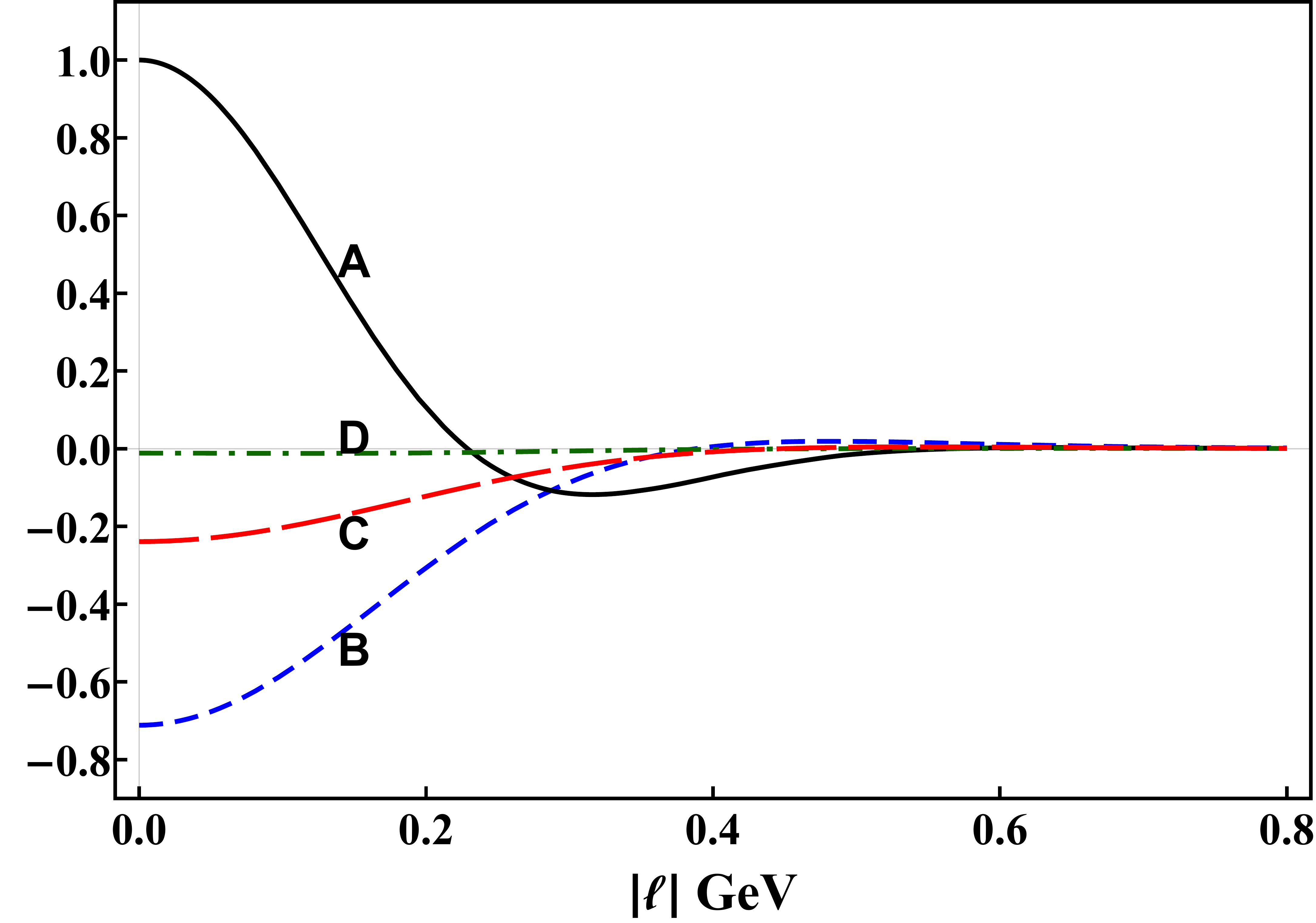}\vspace*{-0ex}
\end{tabular}
\begin{tabular}{lr}
\hspace*{-1ex}\includegraphics[clip,width=0.43\linewidth]{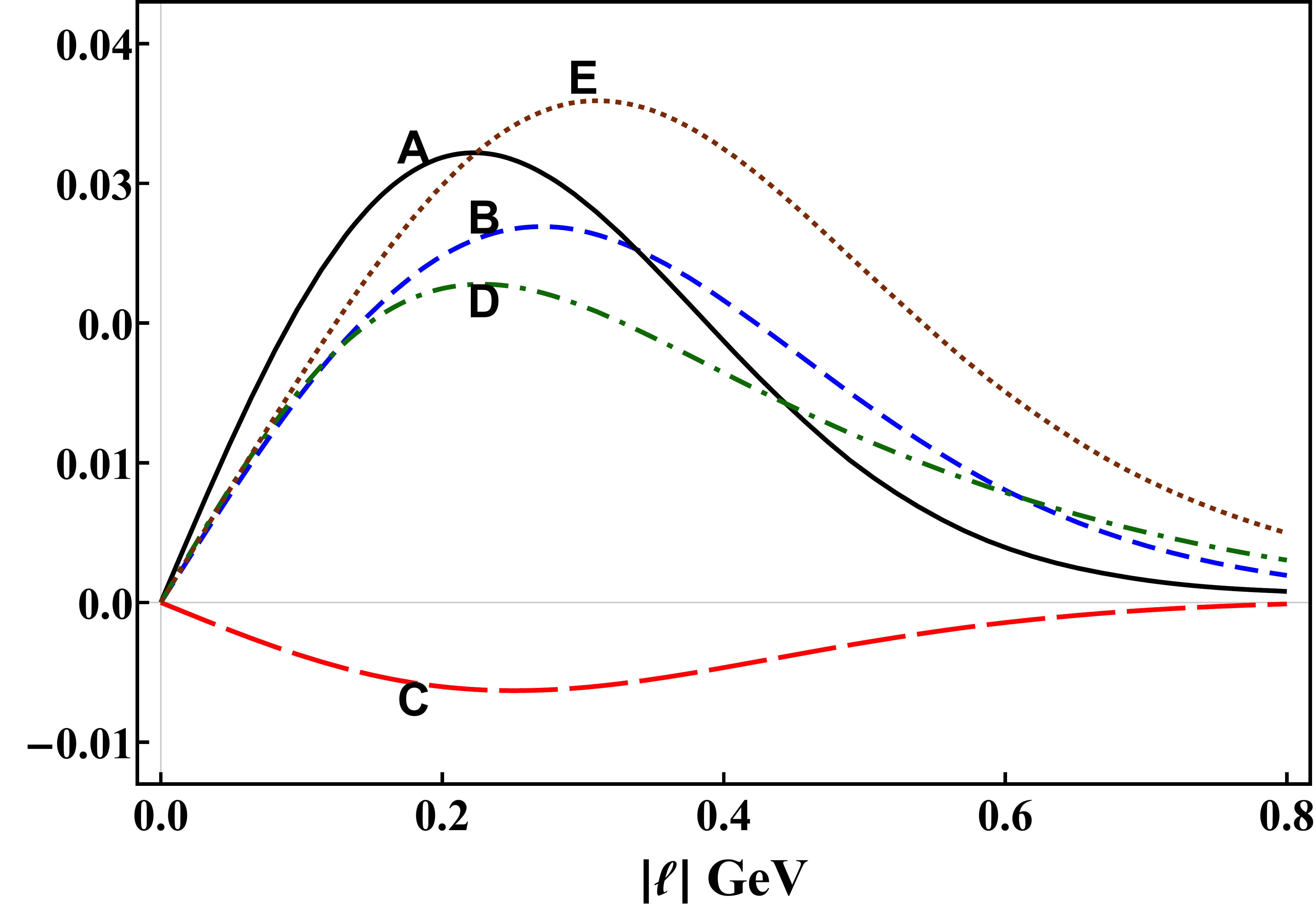}\hspace*{2ex } &
\includegraphics[clip,width=0.43\linewidth]{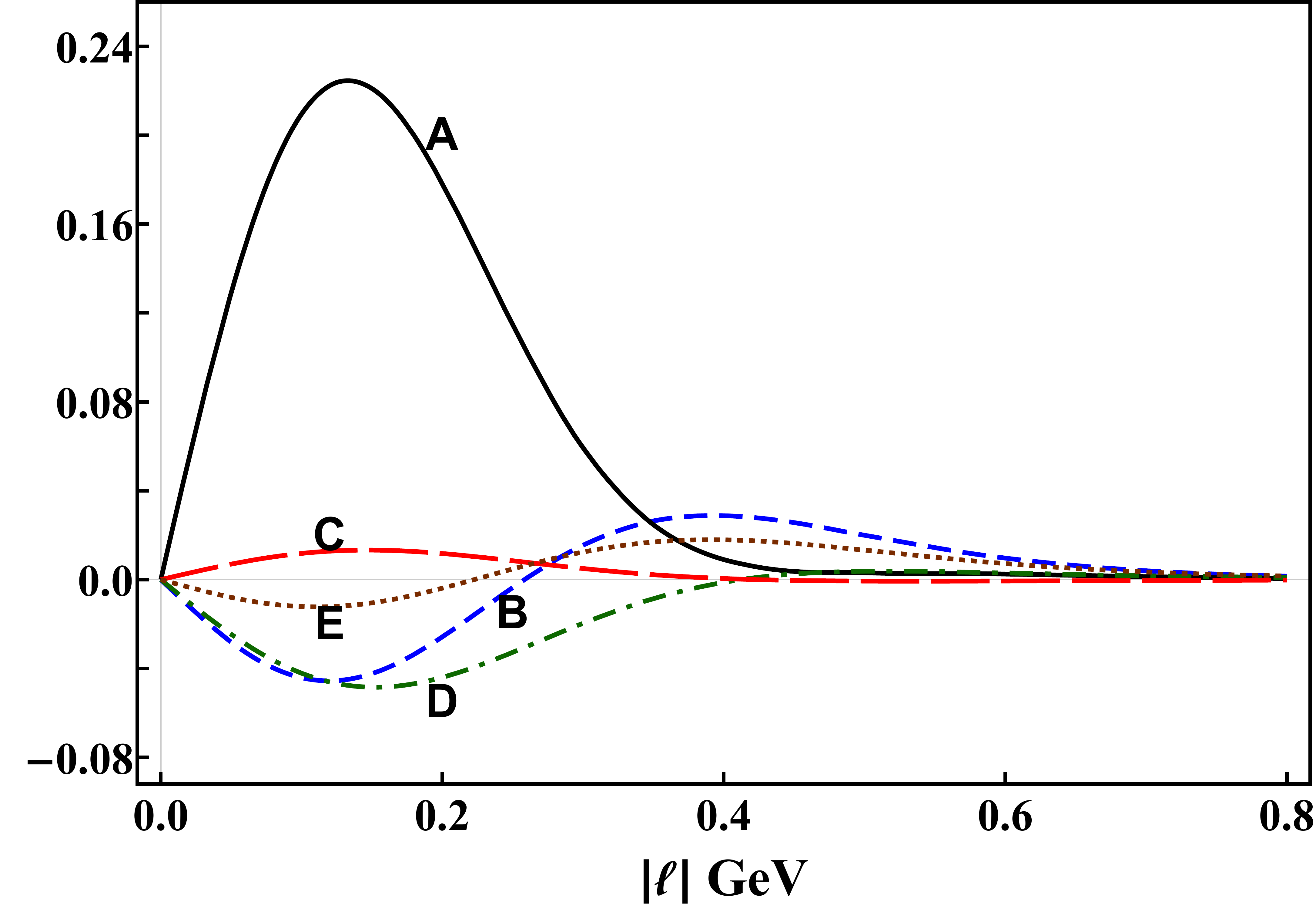}\vspace*{-1ex}
\end{tabular}
\end{center}
\caption{\label{8lambdaS}
$\Lambda$-baryon Faddeev wave functions, Chebyshev-moment projections, Eq.\,\eqref{Wproject}.
\underline{$S$-wave}: $\Lambda_{\rm n=0}$ (\emph{upper-left panel}) and $\Lambda_{\rm n=1}$ (\emph{upper-right}).  Legend.
``A'' $ \to \tilde{\mathpzc s}_{1}^1$;
``B'' $\to \tilde{\mathpzc s}_{1}^{[2,3]}$;
``C'' $\to (\tilde{\mathpzc a}_{3}^{[6,8]}+2\tilde{\mathpzc a}_{5}^{[6,8]})/3$; and
``D'' $\to \tilde{\mathpzc a}_{2}^{[6,8]}$.
\underline{$P$-wave}: $\Lambda_{\rm n=0}$ (\emph{lower-left panel}) and $\Lambda_{\rm n=1}$ (\emph{lower-right}).  Legend.
``A'' $\to \tilde{\mathpzc s}_{2}^1$;
``B'' $\to \tilde{\mathpzc s}_{2}^{[2,3]}$;
``C'' $\to (\tilde{\mathpzc a}_{4}^{[6,8]}+2\tilde{\mathpzc a}_{6}^{[6,8]})/3$;
``D'' $\to \tilde{\mathpzc a}_{1}^{[6,8]}$; and
``E'' $\to (\tilde{\mathpzc a}_{4}^{[6,8]}-\tilde{\mathpzc a}_{6}^{[6,8]})$.
$D$-waves are negligible for all octet baryons considered herein.
}
\end{figure*}

Fig.\,\ref{Loctet} reveals that, concerning the rest-frame quark-di\-quark orbital angular momentum fractions in octet baryons, both measures deliver the same qualitative picture of each baryon's internal structure.  It follows that there is little mixing between partial waves in the computation of a baryon's mass.
Regarding Fig.\,\ref{Ldecuplet}, this is also true for the decuplet states, but there are greater quantitative dissimilarities.  It is nevertheless evident in both panels that $S$-wave strength is shifted into $D$-wave contributions within decuplet positive-parity excitations, as has previously been observed \cite{Eichmann:2016nsu, Qin:2018dqp}.

It is here worth contrasting these results for low-lying positive-parity baryons with those obtained elsewhere \cite{Chen:2017pse} for the two lightest $(I,J^P)=(1/2,1/2^-)$ partners of the nucleon.  For those systems: no solution is obtained unless $P$-waves are present; $P$-waves are the largest component of the rest-frame wave function and dominant in determining the mass, with $S$-waves bringing some attraction; and $D$-waves are negligible.

The remarks closing Sec.\,\ref{SecDiquark} highlight that it is also desirable to explore the effects of various angular momentum components of a given baryon's wave function on observables accessible at modern facilities. This has already been shown to provide valuable insights, \emph{e.g}.\ in connection with the possible appearance of a zero in the proton and neutron elastic form factors \cite{Segovia:2015ufa}, a feature which is greatly influenced by $P$-wave components in the nucleon's rest-frame wave function even though their effect on the nucleon mass is small.

\subsection{Pointwise Structure}
\label{SecPointwise}
The results described hitherto reveal global (integrated) features of the octet and decuplet baryons and their first positive-parity excitations.  It is also worth exposing aspects of their local structure as it is expressed in the pointwise behaviour of their Faddeev amplitudes.  To this end, we consider the zeroth Chebyshev moment of all partial waves in a given baryon's rest-frame Faddeev wave function,  \emph{i.e}.\ projections of the form
\begin{equation}
\label{Wproject}
{\mathpzc Y}(\ell^2;P^2) = \frac{2}{\pi} \int_{-1}^1 \! du\,\sqrt{1-u^2}\,
{\mathpzc Y}(\ell^2,u; P^2)\,,
\end{equation}
where $u=\ell\cdot P/\sqrt{\ell^2 P^2}$.

The order-zero Chebyshev projection of the quark-core Faddeev amplitudes for the nucleon and its positive-parity excitation are plotted in Ref\,\cite{Chen:2017pse}, Fig.\,4; and we reproduce those results.
Herein, therefore, in Fig.\,\ref{8lambdaS} we elect to depict the projections for the $\Lambda$-baryon and its first positive-parity excitation, which are qualitatively equivalent.  Namely, each projection for the $\Lambda_{n=0}$ is of a single sign (positive or negative).  Those associated with the quark core of the $\Lambda_{n=1}$ are quite different: all $S$- and $P$-wave components exhibit a single zero at some point within $0.2< |\ell|/{\rm GeV}\lesssim 0.4$.
Similar statements hold true of the $\Sigma_{n=0,1}$ and $\Xi_{n=0,1}$.
Drawing upon experience with quantum mechanics and with excited-state mesons studied via the Bethe-Salpeter equation \cite{Holl:2004fr, Qin:2011xq, Li:2016dzv, Li:2016mah}, this pattern of behaviour for the first positive-parity excited states indicates that each one may be interpreted as the simplest radial excitation of its ground-state partner.
It is also worth remarking that the relative magnitudes of these Faddeev amplitude projections are consistent with the angular momentum contents indicated by Fig.\,\ref{Loctet}.

As illustrated by the $\Xi^\ast_{n=0,1}$ projections plotted in Fig.\,\ref{10xistarS}, this pattern of behaviour is repeated in decuplet baryons: all zeroth-Chebyshev projections of decuplet ground-state Faddeev wave functions are of a single sign whereas each of these projections for the first positive-parity excitation possesses a single zero.  Once again, therefore, the first positive-parity excitation can be identified as the simplest radial excitation of the ground-state's quark-diquark core.

These observations and conclusions match those in Refs.\,\cite{Segovia:2015hra, Eichmann:2016hgl, Lu:2017cln, Chen:2017pse} and extend them to baryons with strangeness.

\section{Summary and Perspective}
\label{epilogue}
Using a Faddeev kernel that supports a uniformly good description of observed properties of the nucleon, $\Delta$-baryon and Roper resonance, we computed the spectrum and Poincar\'e-covariant wave functions for all flavour-$SU(3)$ octet and decuplet baryons and their first positive-parity excitations (Sec.\,\ref{SecSpectrum}).
A basic prediction of such Faddeev equation studies is the presence of strong nonpointlike, fully-interacting quark-quark (diquark) correlations within all baryons; and our analysis confirms that for a realistic description of these states, it is necessary and sufficient to retain only flavour-$\bar 3$--scalar and flavour-$6$--pseudovector correlations (Sec.\,\ref{SecDiquark}).  Namely, negative-parity diquarks are negligible in these positive-parity baryons.
Moreover, in its rest-frame, every system considered may be judged as primarily $S$-wave in character (Sec.\,\ref{SecJLS}); and the first positive-parity excitation of each octet or decuplet baryon exhibits the characteristics of a radial excitation of the ground-state (Sec.\,\ref{SecPointwise}).

In arriving at these conclusions, we draw a similar picture to quark model descriptions of these systems, so long as rest-frame orbital angular momentum is identified with that existing between dressed-quarks and \mbox{-diquarks}, which are the correct strong-interaction quasiparticle degrees-of-freedom at the hadronic scale and on a material domain extending beyond.
In addition, we confirm the quark-model result that each ground-state octet and decuplet baryon possesses a radial excitation and consequently predict the existence of positive-parity excitations of the $\Xi$, $\Xi^\ast$, $\Omega$ baryons, with the masses listed in Eq.\,\eqref{PredictedBWmasses}.

Our structural predictions for the octet and decuplet baryons and their first positive-parity excitations should be tested because they focus attention on two key questions in baryon structure \cite{Roberts:2016vyn}, \emph{viz}.\ is the expression of emergent mass generation the same in each baryon; and are dynamical quark-quark correlations an essential element in the structure of all baryons?  With the analysis herein, we answer ``yes'' to both questions for the entire array of octet and decuplet baryons and their first positive-parity excitations.

It is likely that for $u, d$-quark baryons, this challenge will be addressed by new generation resonance electroproduction experiments at JLab \cite{Burkert:2018nvj} because, as recent progress toward understanding the Roper resonance has shown \cite{Mokeev:2015lda, Burkert:2017djo, Mokeev:2018zxt}, such experiments probe the quark-cores of the baryons involved when they employ photon virtualities $Q^2 \gtrsim 2\,m_N^2$, \emph{i.e}.\ beyond the meson-cloud domain.  Access to the structure of ground- and excited-state hyperons, on the other hand, may require measurements of hyperon radiative transitions.  One task now, therefore, is to complement studies of the nucleon-to-Roper and nucleon-to-$\Delta$ transitions by delivering predictions for an array of transition form factors involving the baryons described herein on the entire domain of photon virtualities that is accessible at modern facilities.

Another direction is the extension of our analysis to the negative-parity partners of all states considered herein.  This would probably show that the complex structural features of the $(I,J^P)=(1/2,1/2^-)$ states revealed elsewhere \cite{Eichmann:2016nsu, Chen:2017pse} are also expressed in the flavour-$SU(3)$ analogues, \emph{e.g}.\
negative-parity diquark correlations are important;
the Poincar\'e-covariant wave function of each such negative-parity system are predominantly $P$-wave in nature, but possess measurable $S$-wave components;
and the first negative-parity excitation of a given octet or decuplet negative-parity ground-state possesses little of the character of a radial excitation.
However, such expectations should be verified; and empirical consequences elucidated of whatever structural features are uncovered.

Furthermore, with the Faddeev amplitudes of the octet baryons in hand, calculations of the axial couplings and form factors of all these states are within reach, updating and extending earlier analyses of the nucleon \cite{Roberts:2007jh, Eichmann:2011pv, Chang:2012cc}.  The axial couplings of strange baryons have many impacts, being important, \emph{inter alia}, to understanding hypernuclear physics and hence the equation of state for neutron stars \cite{Lonardoni:2014bwa}.  They have recently been calculated using lattice-regularised QCD \cite{Savanur:2018jrb}, thereby enabling potentially valuable theoretical comparisons with future results from our framework which, \emph{e.g}.\ may shed additional light on the role of meson-baryon final-state-interactions.

\acknowledgments
We are grateful: for constructive comments and encouragement from V.\,Burkert, D.~Carman, L.~Chang, \mbox{Z.-F.~Cui}, O.~Denisov, F.~de Soto, K.~Hicks, B.~Grube, A.~Guskov, Y.~Lu,  R.~Gothe, V.~Mokeev, S.-X.~Qin, J.~Rodr\'{\i}guez-Quintero, F.~Wang and S.-S.~Xu;
for the hospitality and support of RWTH Aachen University, III.\,Physikali\-sches Institut B, Aachen - Germany;
and likewise at the University of Huelva, Huelva - Spain, and the University of Pablo de Olavide, Seville - Spain, during the ``4th Workshop on Nonperturbative QCD'' at the University of Pablo de Olavide, 6-9 November 2018.
Work supported by:
Conselho Nacional de Desenvolvimento Cient{\'{\i}}fico e Tecnol{\'o}gico - CNPq, Grant Nos.\,305894/2009-9, 464898/2014-5 (INCT F{\'{\i}}sica Nuclear e Aplica{\c{c}}{\~o}es);
Funda\c{c}\~ao de Amparo \`a Pesquisa do Estado de S\~ao Paulo - FAPESP Grant Nos.\,2013/01907-0,
2015/21550-4;
Jiangsu Province \emph{Hundred Talents Plan for Professionals};
U.S.\ Department of Energy, Office of Science, Office of Nuclear Physics, under contract no.\,DE-AC02-06CH11357;
and
Forschungszentrum J\"ulich GmbH.


\appendix
\setcounter{equation}{0}
\setcounter{figure}{0}
\setcounter{table}{0}
\renewcommand{\theequation}{\Alph{section}.\arabic{equation}}
\renewcommand{\thetable}{\Alph{section}.\arabic{table}}
\renewcommand{\thefigure}{\Alph{section}.\arabic{figure}}

\section{Dressed quark propagator}
\label{appendixKFE}
The dressed-quark propagator can be written:
\begin{subequations}
\begin{align}
S_f(p) & =  -i \gamma\cdot p\, \sigma_V^f(p^2) + \sigma_S^f(p^2) \\
& = 1/[i\gamma\cdot p\, A_f(p^2) + B_f(p^2)]\,.
\label{SpAB}
\end{align}
\end{subequations}
It is known that for light-quarks the wave function renormalisations and dressed-quark masses:
\begin{equation}
\label{ZMdef}
Z_f(p^2)=1/A_f(p^2)\,,\;M_f(p^2)=B_f(p^2)/A_f(p^2)\,,
\end{equation}
respectively, receive strong momentum-dependent corrections at infrared momenta \cite{Lane:1974he, Politzer:1976tv, Zhang:2004gv, Bhagwat:2004kj, Bhagwat:2006tu, Binosi:2016wcx}: $Z_f(p^2)$ is suppressed and $M_f(p^2)$ enhanced.  These features are an expression of DCSB and, plausibly, of confinement \cite{Horn:2016rip}; and their impact on hadron phenomena has long been emphasised \cite{Roberts:1994hh}.

Numerical solutions of the quark gap equation are now readily obtained \cite{Binosi:2016wcx}.  However, the utility of an algebraic form for $S_f(p)$ when calculations require the evaluation of numerous multidimensional integrals is self-evident.  An efficacious parametrisation of $S_f(p)$, which exhibits the features described above, is expressed via \cite{Ivanov:1998ms, Ivanov:2007cw}:
{\allowdisplaybreaks
\begin{subequations}
\label{EqSSSV}
\begin{align}
\bar\sigma_S^f(x) & =  2\,\bar m_f \,{\cal F}(2 (x+\bar m_f^2)) \nonumber \\
& \quad + {\cal F}(b_1^f x) \,{\cal F}(b_3^f x) \,
\left[b_0^f + b_2^f {\cal F}(\epsilon x)\right]\,,\label{ssm} \\
\label{svm} \bar\sigma_V^f(x) & =  \frac{1}{x+\bar m_f^2}\, \left[ 1 - {\cal F}(2
(x+\bar m_f^2))\right]\,,
\end{align}
\end{subequations}}
\hspace*{-0.5\parindent}with $x=p^2/\lambda^2$, $\bar m_f$ = $m_f/\lambda$,
\begin{equation}
\label{defcalF}
{\cal F}(x)= \frac{1-\mbox{\rm e}^{-x}}{x}  \,,
\end{equation}
$\bar\sigma_S(x) = \lambda\,\sigma_S(p^2)$, $\bar\sigma_V(x) =
\lambda^2\,\sigma_V(p^2)$.
The mass-scale,
\begin{equation}
\label{lambdaval}
\lambda=0.566\,{\rm GeV},
\end{equation}
and parameter values
\begin{equation}
\label{tableA}
\begin{array}{l|ccccc}
   & \bar m_f& b_0^f & b_1^f & b_2^f & b_3^f \\\hline
   u & 0.00897 & 0.131 & 2.90 & 0.603 & 0.185 \\
   s & 0.210\phantom{97} & 0.105 & 3.18 & 0.858 & 0.185
\end{array}\;,
\end{equation}
associated with Eqs.\,\eqref{EqSSSV} were fixed in a fit to wide range of meson observables \cite{Ivanov:1998ms, Ivanov:2007cw}.  (These values should be understood as determined at a renormalisation scale $\zeta=1\,$GeV; and $\epsilon=10^{-4}$ in Eq.\ (\ref{ssm}) acts only to decouple the large- and intermediate-$p^2$ domains.)

The dimensionless $u=d$ and $s$ current-quark masses in Eq.\,(\ref{tableA}) correspond to
\begin{equation}
\label{mcq}
m_u=5.08\,{\rm MeV}, \; 
m_s = 119\,{\rm MeV},
\end{equation}
and the parametrisation yields the following Euclidean constituent-quark masses, defined as the solution $M^E = p$, $p^2=M^2(p^2)$:
\begin{equation}
\label{MEq}
M_{u,d}^E = 0.33\,{\rm GeV},\;
M_{s}^E = 0.49\,{\rm GeV}.
\end{equation}
The ratio $M_u^E/m_u = 65$ is one expression of DCSB in the parametrisation of $S_u(p)$.  It emphasises the dramatic enhancement of the dressed light-quark mass function at infrared momenta.  On the other hand, the result $M_s^E/m_s = 4.2$ highlights once again that the $s$-quark lies close to the transition boundary between dominance of emergent mass generation over that connected with the Higgs boson \cite{Ding:2015rkn, Ding:2018xwy}.

As with the diquark propagators in Eq.\,\eqref{Eqqqprop}, the expressions in Eq.\,\eqref{EqSSSV} ensure confinement of the dressed quarks via the violation of reflection positivity.

\section{Assorted Formulae}
\label{appendixAssorted}
As noted in Sec.\,\ref{SecDressedq}, we assume isospin symmetry throughout, in which case it suffices to specify the following spin-flavour column vectors for the octet baryon Faddeev amplitudes:
{\allowdisplaybreaks
\begin{subequations}
\label{FaddeevSpinFlavour}
\begin{align}
u_p  = & \left[\begin{array}{c}
u[ud]_{0^+} \\ d\{uu\}_{1^+} \\ u\{ud\}_{1^+}
\end{array}\right]
\leftrightarrow\left[\begin{array}{c}
\mathpzc{s}^1_p \\ \mathpzc{a}^4_p \\ \mathpzc{a}^5_p
\end{array}\right]\,,
\\
u_\Lambda =\frac{1}{\sqrt{2}} & \left[\begin{array}{c}
\surd 2\, s[ud]_{0^+} \\ d[us]_{0^+}-u[ds]_{0^+} \\ d\{us\}_{1^+}-u\{ds\}_{1^+} \end{array}\right]
\leftrightarrow\left[\begin{array}{c}
\mathpzc{s}^1_\Lambda \\ \mathpzc{s}^{[2,3]}_\Lambda \\ \mathpzc{a}^{[6,8]}_\Lambda \end{array}\right],\label{flavourLambda}\\
u_\Sigma = & \left[\begin{array}{c}
u[us]_{0^+} \\ s\{uu\}_{1^+} \\ u\{us\}_{1^+} \end{array}\right]
\leftrightarrow\left[\begin{array}{c}
\mathpzc{s}^2_\Sigma \\ \mathpzc{a}^4_\Sigma \\ \mathpzc{a}^6_\Sigma  \end{array}\right]\,, \label{flavourSigma}
\\
u_\Xi = & \left[\begin{array}{c}
s[us]_{0^+} \\ s\{us\}_{1^+} \\ u\{ss\}_{1^+}  \end{array}\right]
\leftrightarrow\left[\begin{array}{c}
\mathpzc{s}^2_\Sigma \\ \mathpzc{a}^6_\Sigma \\ \mathpzc{a}^9_\Sigma \end{array}\right]\,,
\end{align}
\end{subequations}}
\hspace*{-0.5\parindent}%
where $[\cdot]_{J^P}$ and $\{\cdot \}_{J^P}$ denote, respectively, flavour combinations generated by $T^{j=1,2,3}_{\bar 3_f}$ and $T^{g=1,\dots,6}_{6_f}$ in Eqs.\,\eqref{Tmatrices}, with the subscript indicating the spin-parity of the associated correlation.  Naturally, the same vector applies to both ground-states and their radial excitations.
The difference between these states is expressed in the values of the coefficients $\mathpzc{s}^j_{\underline{8} k(=1,2)}$ and $\mathpzc{a}^g_{\underline{8} k(=1,\dots,6)}$ that appear in Eq.\,\eqref{8sa} and are obtained by solving the appropriate Faddeev equations.  A shorthand notation for these coefficients, which expresses their connection with Eqs.\,\eqref{Tmatrices}, is specified by the rightmost column of each of Eqs.\,\eqref{FaddeevSpinFlavour}: superscript ``1'' connects with $T^1_{\bar 3_f}$, \ldots\,, superscript ``4''\,$\to T^1_{6_f}$, \ldots, superscript ``9''\,$\to T^6_{6_f}$.

It is worth comparing Eqs.\,\eqref{flavourLambda} and \eqref{flavourSigma}, with the latter adapted to the neutral $\Sigma^0$ case following Eq.\,(49) in Ref.\,\cite{Chen:2012qr}.  Whilst the $\Lambda^0$ and $\Sigma^0$ baryons are associated with the same combination of valence-quarks,
their spin-flavour wave functions are different: the $\Lambda_{I=0}^0$ contains more of the lighter $J=0$ diquark correlations than the $\Sigma_{I=1}^0$.  It follows that the $\Lambda^0$ must be lighter than the $\Sigma^0$.  

The analogous vectors for the decuplet baryons are:
{\allowdisplaybreaks
\begin{subequations}
\label{DecupletSpinFlavour}
\begin{align}
u_\Delta = & \left[ u \{uu\}_{1^+} \right] \leftrightarrow [\mathpzc{a}^4_\Delta]\,,\\
u_{\Sigma^\ast} = & \left[\begin{array}{c}
s\{uu\}_{1^+}\\ u \{us\}_{1^+}
\end{array}\right] \leftrightarrow
\left[\begin{array}{c}
\mathpzc{a}^4_{\Sigma^\ast} \\ \mathpzc{a}^6_{\Sigma^\ast}
\end{array}\right]\,,\\
u_{\Xi^\ast} = & \left[\begin{array}{c}
s\{us\}_{1^+}\\ u \{ss\}_{1^+}
\end{array}\right] \leftrightarrow
\left[\begin{array}{c}
\mathpzc{a}^6_{\Xi^\ast} \\ \mathpzc{a}^9_{\Xi^\ast}
\end{array}\right]\,,\\
u_\Omega = & \left[ s \{ss\}_{1^+} \right] \leftrightarrow [\mathpzc{a}^9_\Omega]\,.
\end{align}
\end{subequations}}

Naturally, owing to isospin symmetry, it is only necessary to explicitly consider one state in any isospin multiplet: all members of the multiplet are degenerate; and the flavour structures of the Faddeev amplitudes are simply related by isospin raising or lowering operations.  For instance, in the $\Delta$-baryon quadruplet, one has:
{\allowdisplaybreaks
\begin{subequations}
\begin{align}
u_{\Delta^{++}}= u_\Delta = & \left[ u \{uu\}_{1^+} \right]\,,\\
u_{\Delta^{+}} = & \left[\begin{array}{c}
u\{ud\}_{1^+} \sqrt{\frac{2}{3}}\\
d\{uu\}_{1^+} \sqrt{\frac{1}{3}}
\end{array}\right]\,,\\
u_{\Delta^{0}} = & \left[\begin{array}{c}
d\{ud\}_{1^+} \sqrt{\frac{2}{3}}\\
u\{dd\}_{1^+} \sqrt{\frac{1}{3}}
\end{array}\right]\,,\\
u_{\Delta^{-}}= & \left[ d \{dd\}_{1^+} \right]\,.
\end{align}
\end{subequations}}

\addtocounter{section}{1}

\begin{table*}[!t]
\caption{\label{FaddeevAmps}
Diquark fractions within octet and decuplet baryons and their first radial excitations, measured by their contribution to the associated Faddeev amplitude: see Eqs.\,\eqref{mathbbD}, \eqref{Dfracs}.
The superscript on ${\mathpzc s}^j$, ${\mathpzc a}^g$ is a diquark enumeration label: see Eqs.\,\protect\eqref{FaddeevSpinFlavour}.  In reading the decuplet results, recall:
$a^4 \leftrightarrow \{uu\}_{1^+}$,
$a^6 \leftrightarrow \{us\}_{1^+}$,
$a^9 \leftrightarrow \{ss\}_{1^+}$.
The rightmost column lists the net scalar diquark fractions, \emph{i.e}.\ the probability, by this measure, of finding a scalar diquark in the identified baryon.
}
\begin{center}
\begin{tabular*}
{\hsize}
{
l@{\extracolsep{0ptplus1fil}}
l@{\extracolsep{0ptplus1fil}}
|r@{\extracolsep{0ptplus1fil}}
r@{\extracolsep{0ptplus1fil}}
r@{\extracolsep{0ptplus1fil}}
|r@{\extracolsep{0ptplus1fil}}
r@{\extracolsep{0ptplus1fil}}
r@{\extracolsep{0ptplus1fil}}
r@{\extracolsep{0ptplus1fil}}
r@{\extracolsep{0ptplus1fil}}
|r@{\extracolsep{0ptplus1fil}}
}\hline
& & ${\mathpzc s}^1$ & ${\mathpzc s}^2$ & ${\mathpzc s}^{[2,3]}\ $ &
    ${\mathpzc a}^4$ & ${\mathpzc a}^5$ & ${\mathpzc a}^6$ &
    ${\mathpzc a}^{[6,8]}$ & ${\mathpzc a}^9\ $ &
   $P_{J^P=0^+} $ \\\hline
Octet ($n=0$)
& $N$ & 0.37 & & & 0.42 & 0.21 & & & & 37\%\\
& $\Lambda$ & 0.21 & & 0.19$\ $ & & & & 0.60 & & 40\%\\
& $\Sigma$ & & 0.47 & & 0.34 & & 0.19 & & & 47\%\\
& $\Xi$ & & 0.51 & & & & 0.22 & & 0.27$\ $ & 51\%\\\hline
Octet ($n=1$)
& $N$ & 0.39 & & & 0.41 & 0.20 & & & & 39\%\\
& $\Lambda$ & 0.16 & & 0.38$\ $ & & & & 0.46 & & 54\%\\
& $\Sigma$ & & 0.41 & & 0.26 & & 0.34 & & & 41\%\\
& $\Xi$ & & 0.47 & & & & 0.23 & & 0.30$\ $ & 47\%\\\hline
Decuplet ($n=0$)
& $\Delta$ & & & & 1.0$\phantom{7}$ & & & & & 0\%\\
& $\Sigma^\ast$ & & & & 0.27 & & 0.73 & & & 0\%\\
& $\Xi^\ast$ & & & & & & 0.72 & & 0.28$\ $ & 0\%\\
& $\Omega$ & & & & & & & & 1.0$\phantom{7}$$\ $ & 0\%\\\hline
Decuplet ($n=1$)
& $\Delta$ & & & & 1.0$\phantom{7}$ & & & & & 0\%\\
& $\Sigma^\ast$ & & & & 0.27 & & 0.73 & & & 0\%\\
& $\Xi^\ast$ & & & & & & 0.65 & & 0.35$\ $ & 0\%\\
& $\Omega$ & & & & & & & & 1.0$\phantom{7}$$\ $ & 0\%\\\hline
\end{tabular*}
\end{center}
\end{table*}

\begin{table*}[!t]
\caption{\label{qqmasses}
Diquark fractions within octet and decuplet baryons and their first radial excitations, measured by their contribution to the associated mass: see Eqs.\,\eqref{qqPmasses}.
The superscript on ${\mathpzc s}^j$, ${\mathpzc a}^g$ is a diquark enumeration label: see Eqs.\,\protect\eqref{FaddeevSpinFlavour}.  In reading the decuplet results, recall:
$a^4 \leftrightarrow \{uu\}_{1^+}$,
$a^6 \leftrightarrow \{us\}_{1^+}$,
$a^9 \leftrightarrow \{ss\}_{1^+}$.
The rightmost column lists the net scalar diquark fractions, \emph{i.e}.\ the probability, by this measure, of finding a scalar diquark in the identified baryon.
}
\begin{center}
\begin{tabular*}
{\hsize}
{
l@{\extracolsep{0ptplus1fil}}
l@{\extracolsep{0ptplus1fil}}
|r@{\extracolsep{0ptplus1fil}}
r@{\extracolsep{0ptplus1fil}}
r@{\extracolsep{0ptplus1fil}}
|r@{\extracolsep{0ptplus1fil}}
r@{\extracolsep{0ptplus1fil}}
r@{\extracolsep{0ptplus1fil}}
r@{\extracolsep{0ptplus1fil}}
r@{\extracolsep{0ptplus1fil}}
|r@{\extracolsep{0ptplus1fil}}
}\hline
& & ${\mathpzc s}^1$ & ${\mathpzc s}^2$ & ${\mathpzc s}^{[2,3]}\ $ &
    ${\mathpzc a}^4$ & ${\mathpzc a}^5$ & ${\mathpzc a}^6$ &
    ${\mathpzc a}^{[6,8]}$ & ${\mathpzc a}^9\ $ &
   $P_{J^P=0^+} $ \\\hline
Octet ($n=0$)
& $N$ & 0.84 & & & 0.09 & 0.07 & & & & 84\%\\
& $\Lambda$ & 0.17 & & 0.76$\ $ & & & & 0.07 & & 93\%\\
& $\Sigma$ & & 0.85 & & 0.10 & & 0.05 & & & 85\%\\
& $\Xi$ & & 0.92 & & & & 0.05 & & 0.03$\ $ & 92\%\\\hline
Octet ($n=1$)
& $N$ & 0.93 & & & 0.04 & 0.03 & & & & 93\%\\
& $\Lambda$ & 0.08 & & 0.88$\ $ & & & & 0.04 & & 96\%\\
& $\Sigma$ & & 0.93 & & 0.05 & & 0.02$\ $ & & & 93\%\\
& $\Xi$ & & 0.96 & & & & 0.02 & & 0.02 & 96\%\\\hline
Decuplet ($n=0$)
& $\Delta$ & & & & 1.0$\phantom{7}$ & & & & & 0\%\\
& $\Sigma^\ast$ & & & & 0.12 & & 0.88 & & & 0\%\\
& $\Xi^\ast$ & & & & & & 0.97 & & 0.03$\ $ & 0\%\\
& $\Omega$ & & & & & & & & 1.0$\phantom{7}$$\ $ & 0\%\\\hline
Decuplet ($n=1$)
& $\Delta$ & & & & 1.0$\phantom{7}$ & & & & & 0\%\\
& $\Sigma^\ast$ & & & & 0.07 & & 0.93 & & & 0\%\\
& $\Xi^\ast$ & & & & & & 0.98 & & 0.02$\ $ & 0\%\\
& $\Omega$ & & & & & & & & 1.0$\phantom{7}$$\ $ & 0\%\\\hline
\end{tabular*}
\end{center}
\end{table*}

\begin{figure*}[!t]
\begin{center}
\begin{tabular}{lr}
\includegraphics[clip,width=0.42\linewidth]{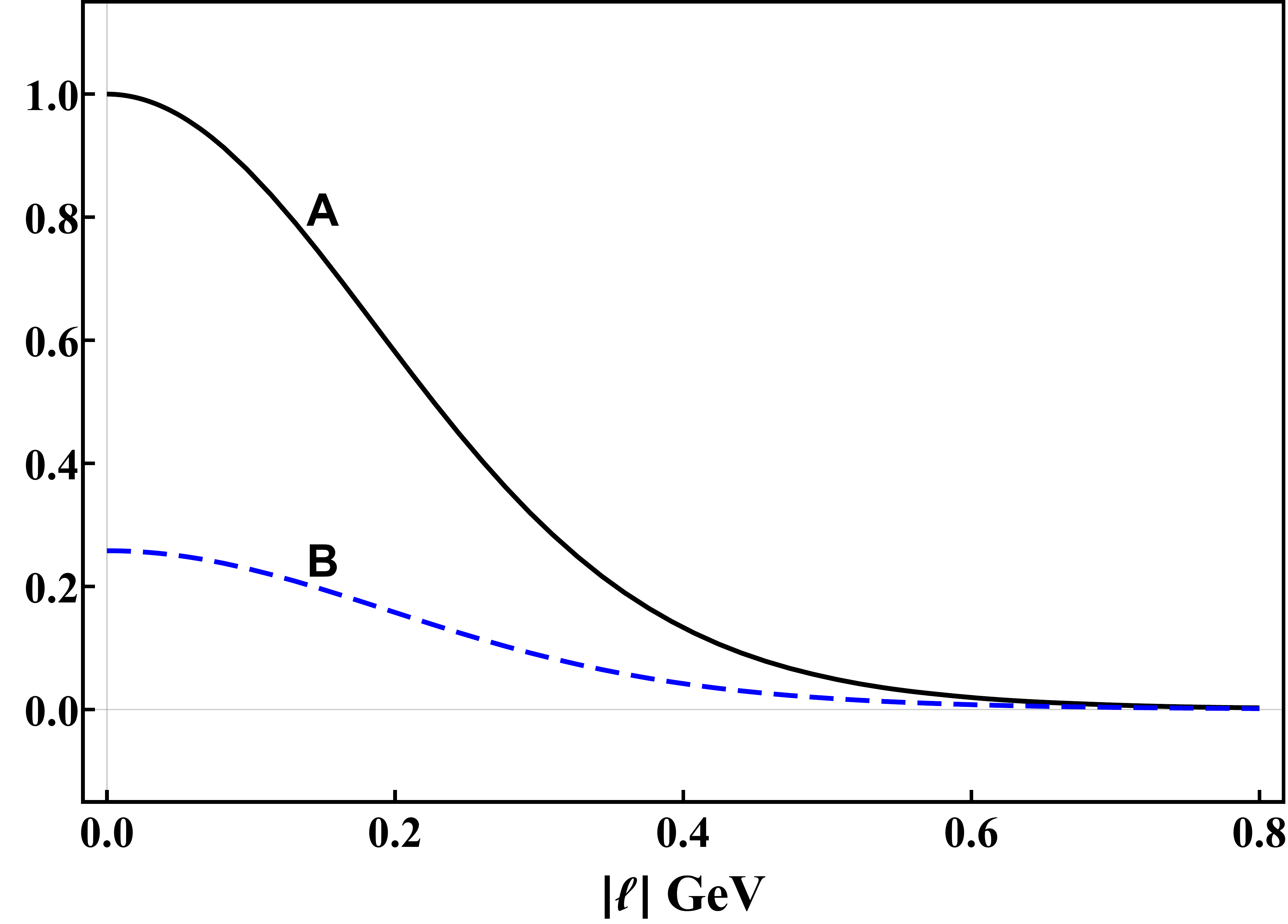}\hspace*{2ex } &
\includegraphics[clip,width=0.42\linewidth]{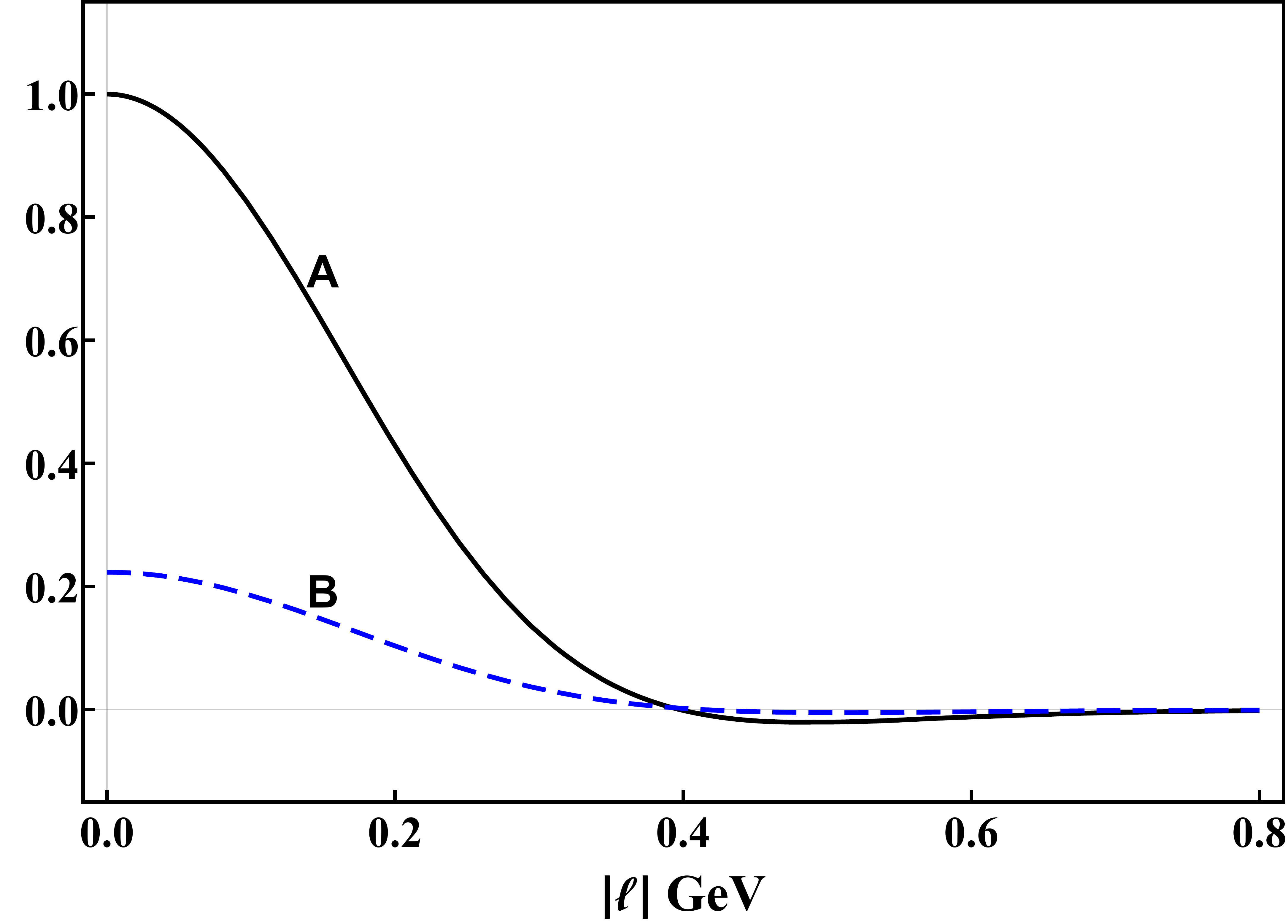}\vspace*{-0ex}
\end{tabular}
\begin{tabular}{lr}
\hspace*{-1.5ex}\includegraphics[clip,width=0.44\linewidth]{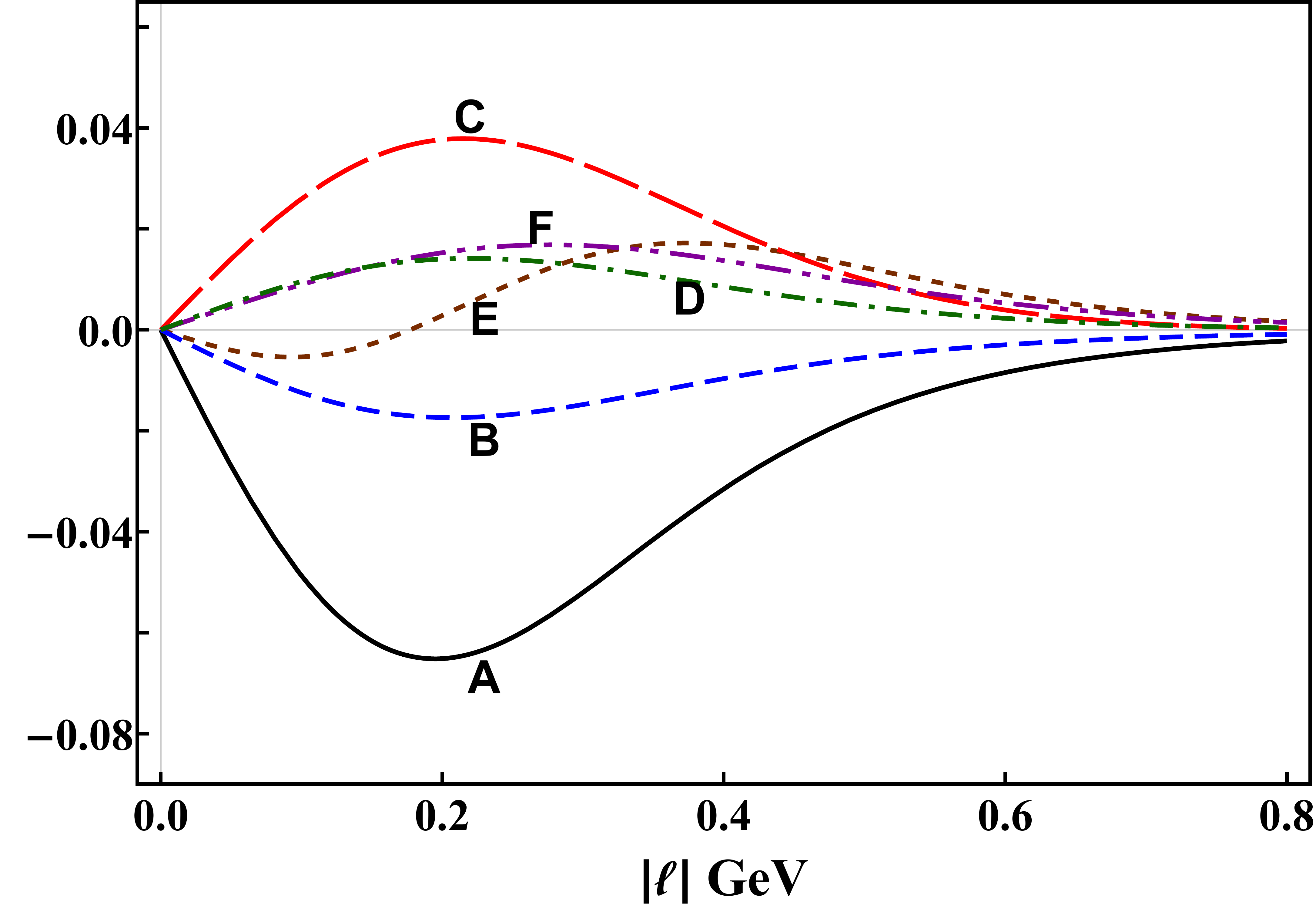}\hspace*{0ex} &
\includegraphics[clip,width=0.44\linewidth]{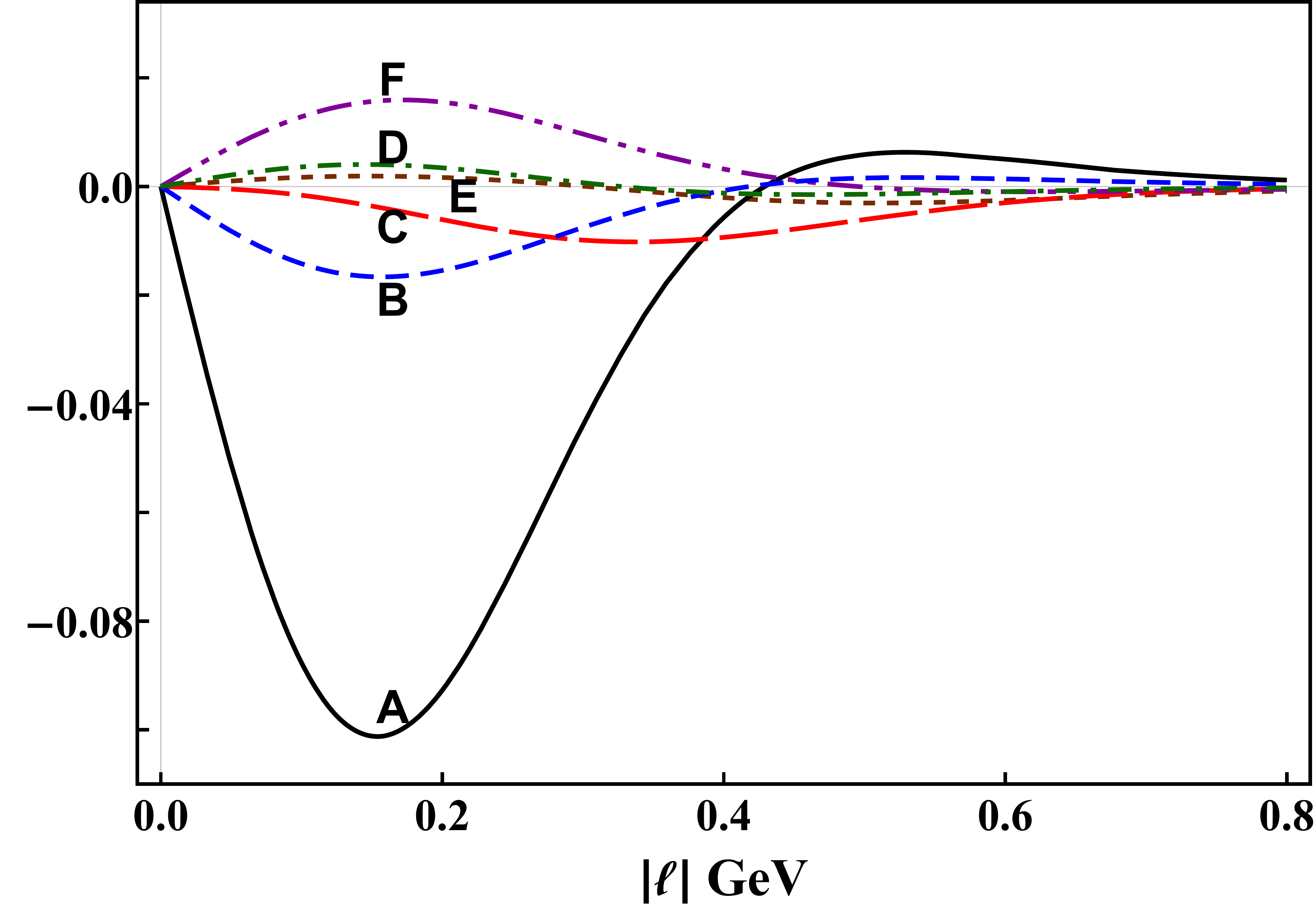}\vspace*{-0ex}
\end{tabular}
\begin{tabular}{lr}
\hspace*{-1ex}\includegraphics[clip,width=0.44\linewidth]{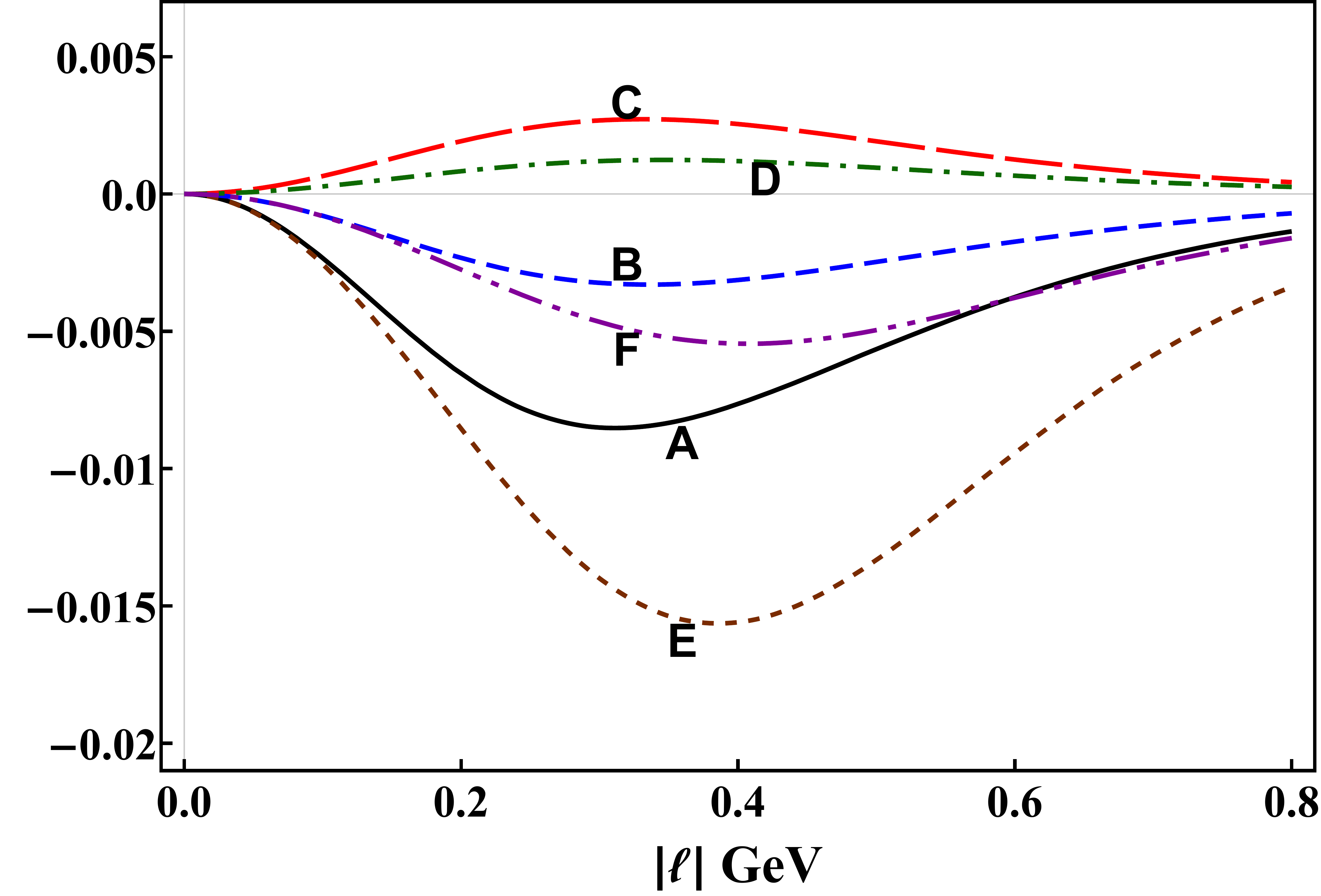}\hspace*{2ex } &
\includegraphics[clip,width=0.42\linewidth]{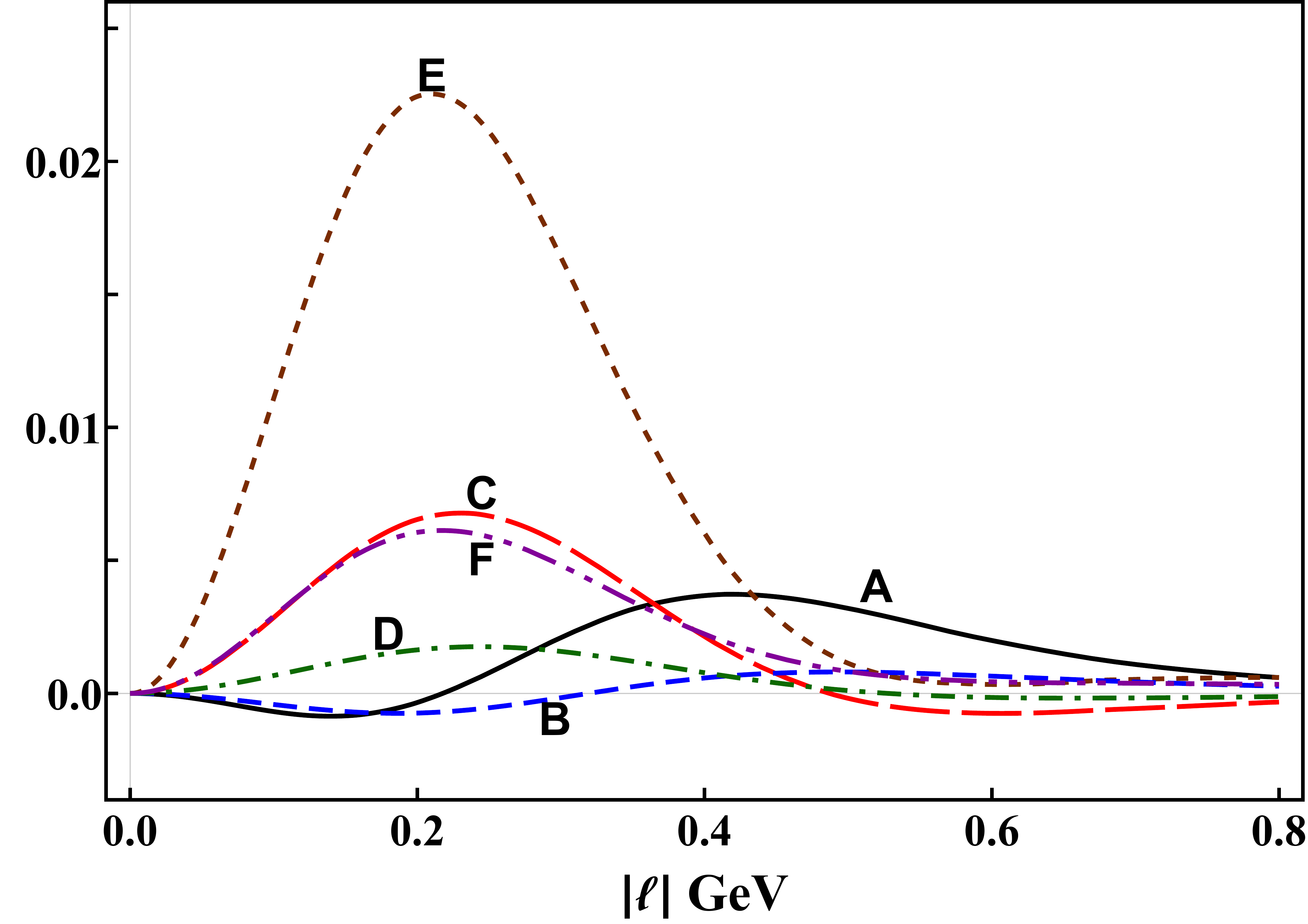}\vspace*{-0ex}
\end{tabular}
\end{center}
\caption{\label{10xistarS}
$\Xi^\ast$-baryon Faddeev wave functions, Chebyshev-moment projections, Eq.\,\eqref{Wproject}.
\underline{$S$-wave}: $\Xi^\ast_{\rm n=0}$ (\emph{top-left panel}) and $\Xi^\ast_{\rm n=1}$ (\emph{top-right}).  Legend.
``A'' $ \to \tilde{\mathpzc a}_{1}^6+(-\tilde{\mathpzc a}_{6}^6+\tilde{\mathpzc a}_{8}^6)/3$; and
``B'' $\to \tilde{\mathpzc a}_{1}^9+(-\tilde{\mathpzc a}_{6}^9+\tilde{\mathpzc a}_{8}^9)/3$.
\underline{$P$-wave}: $\Xi^\ast_{\rm n=0}$ (\emph{middle-left panel}) and $\Xi^\ast_{\rm n=1}$ (\emph{middle-right}).  Legend.
``A'' $\to \tilde{\mathpzc a}_{4}^6$;  
``B'' $\to \tilde{\mathpzc a}_{4}^9$;  
``C'' $\to (2\tilde{\mathpzc a}_{2}^6-\tilde{\mathpzc a}_{5}^6-2\tilde{\mathpzc a}_{7}^6)/3$; 
``D'' $\to (2\tilde{\mathpzc a}_{2}^9-\tilde{\mathpzc a}_{5}^9-2\tilde{\mathpzc a}_{7}^9)/3$; 
``E'' $\to \tilde{\mathpzc a}_{2}^6-(\tilde{\mathpzc a}_{5}^6-\tilde{\mathpzc a}_{7}^6)/5$; and 
``F'' $\to \tilde{\mathpzc a}_{2}^9-(\tilde{\mathpzc a}_{5}^9-\tilde{\mathpzc a}_{7}^9)/5$;  
\underline{$D$-wave}: $\Xi^\ast_{\rm n=0}$ (\emph{bottom-left panel}) and $\Xi^\ast_{\rm n=1} $(\emph{bottom-right}).  Legend.
``A'' $\to\tilde{\mathpzc a}_{3}^6$; 
``B'' $\to\tilde{\mathpzc a}_{3}^9$; 
``C'' $\to-(\tilde{\mathpzc a}_{6}^6+2\tilde{\mathpzc a}_{8}^6)/3$; 
``D'' $\to-(\tilde{\mathpzc a}_{6}^9+2\tilde{\mathpzc a}_{8}^9)/3$;  
``E'' $\to-\tilde{\mathpzc a}_{6}^6+\tilde{\mathpzc a}_{8}^6$; and  
``F'' $\to-\tilde{\mathpzc a}_{6}^9+\tilde{\mathpzc a}_{8}^9$.          
$F$-wave components are negligible for all decuplet baryons considered herein.
}
\end{figure*}

\addtocounter{section}{-1}

\section{Supplementary Tables and Figure}
\label{appendixAssortedTabsFigs}
Here we collect two tables and a figure used above in elucidating structural features of octet and decuplet baryons and their first positive-parity excitations.

In considering the diquark content of decuplet baryons reported in Tables~\ref{FaddeevAmps}, \ref{qqmasses}, the mixed-flavour diquark, when present, is favoured for an obvious reason, \emph{viz}.\ considering the Faddeev equation kernel, it is fed by twice as many exchange processes as the like-flavour correlation.


\end{document}